\pdfoutput=1
\documentclass[prd,11pt,aps,superscriptaddress,floatfix,showkeys,showpacs]{revtex4-1}
\usepackage{xcolor,epsfig,amsmath,amssymb,subfigure,placeins}
\usepackage{graphicx,amsmath,bbm,bm,array}
\usepackage[colorlinks=true,linktocpage=true,linkcolor=red,citecolor=blue]{hyperref}
\def\nn{\nonumber\\}
\def\({\left(}
\def\){\right)}
\def\lL{\ln\frac{\Lambda}{4\pi T}}
\def\be{\begin{eqnarray}}
\def\ee{\end{eqnarray}}
\def\para{\parallel}
\def\hmu{\hat\mu}
\def\hr{\hat r}

\begin{document}
\title{Heavy quarkonium moving in hot and dense deconfined nuclear matter}
\author{Lata Thakur}
\email{latathakur@prl.res.in}
\affiliation{Theory Division, Physical Research Laboratory, Navrangpura, 
Ahmedabad 380 009, India}

\author{Najmul Haque}
\email{Najmul.Haque@theo.physik.uni-giessen.de}
\affiliation{Institut f\"ur Theoretische Physik, Justus-Liebig--Universit\"at 
Giessen,\\ 35392 Giessen, Germany}

\author{Hiranmaya Mishra}
\email{hm@prl.res.in}
\affiliation{Theory Division, Physical Research Laboratory, Navrangpura, 
Ahmedabad 380 009, India}

\begin{abstract}
We study the behavior of the complex potential between a heavy quark and its antiquark, 
which are in relative motion with respect to a hot and dense medium. The heavy 
quark-antiquark complex potential is obtained by correcting both the
Coulombic and the linear terms in the Cornell potential through a dielectric
function estimated within the real-time formalism using the hard thermal loop approximation.
We show the variation of both the real and the imaginary parts of the potential for different 
values of velocities when the bound state ($ Q\bar{Q}$ pair) is aligned in the direction parallel
as well as perpendicular to the relative velocity of the $ Q\bar{Q}$ pair
with the thermal medium. With an increase of the relative 
velocity the screening of the real part of the potential becomes weaker at short distances and 
stronger at large distances for the parallel case. However, for the perpendicular case the potential
decreases with an increase of the velocity at all distances which results in the larger screening of the
potential. In addition, the inclusion of the string term makes the screening of the potential weaker as 
compared to the Coulombic term alone for both cases. Therefore, by combining all these effects we 
expect a stronger binding of a $ Q\bar{Q} $ pair in a moving medium in the presence of the string
term as compared to the Coulombic term alone. The imaginary part decreases (in magnitude) with an
increase of the relative velocity, 
leading to a decrease of the width of the quarkonium state at higher velocities. The inclusion of 
the string term increases the magnitude of the imaginary part, which results in an increase of the
width of the quarkonium states.
All of these effects lead to the modification in the dissolution of quarkonium states.
\end{abstract}



\maketitle

\section{Introduction}
The heavy quarks (HQs) produced in the early stage of the relativistic heavy-ion collisions are important for investigating the properties of the quark gluon plasma 
(QGP). HQs can travel through the thermalized QGP medium, and they can retain the 
information about their interactions with the medium. 
In the QGP medium, plasma screening effects
are expected to modify the heavy quark-antiquark ($ Q\bar{Q} $) potential, and 
above a certain threshold temperature this may lead to a dissolution of $ 
Q\bar{Q} $ bound states~\cite{Matsui:1986dk,Karsch:1987pv}.
As the low-lying heavy quarkonium $Q\bar{Q} $ bound states are the only hadronic states that can survive
in the deconfined phase~\cite{Rapp:2008tf,Mocsy:2013syh}, they are considered as the most 
powerful probes.
Quarkonium suppression can be used to test the formation of a QGP in 
ultrarelativistic heavy-ion collisions. In recent years the quarkonium spectral 
functions and meson current correlators have been studied
using both analytic potential models~\cite{Wong:2004zr,Mocsy:2005qw,Mocsy:2007yj,Mocsy:2007jz,Alberico:2007rg,Mocsy:2004bv,
Karsch:2000gi,Cabrera:2006wh} and numerical lattice QCD models~\cite{Umeda:2002vr,Asakawa:2003re,Datta:2003ww,Jakovac:2006sf,Aarts:2007pk,Ding:2012sp,Ohno:2011zc,
Ding:2012ar,Kaczmarek:2012ne,deForcrand:2000akx,Aarts:2012ka,Aarts:2011sm,Aarts:2013kaa,Aarts:2014cda}.
Additionally, the theoretical study of quarkonia in a thermal medium was also formulated in the language of effective field theory (EFT) where a sequence of 
EFTs~\cite{Brambilla:1999xf,Brambilla:2004jw,Caswell:1985ui,Thacker:1990bm,Bodwin:1994jh,Brambilla:2008cx} are derived by considering a small velocity for the heavy quark, which produces a hierarchy of scales of the heavy-quark bound state: $m_{Q}v^2\ll m_{Q}v\ll m_{Q} $. Moreover, numerical lattice QCD studies the quarkonium suppression by calculating spectral
functions, which are reconstructed from the Euclidean correlators constructed on the 
lattice. However, lattice QCD 
cannot be used directly to reconstruct the spectral functions 
from the quarkonium correlators as it is difficult to perform an
analytic continuation from imaginary to real time. 
Another important theoretical finding in this context was the appearance of
a complex potential at finite temperature~\cite{Laine:2006ns}, which has further stimulated the imaginary potential 
model studies~\cite{Petreczky:2010tk,Margotta:2011ta} as well as attempts to obtain the imaginary potential from lattice QCD~\cite{Rothkopf:2011db,Burnier:2012az}.
It was previously thought that {$Q\bar Q$} states dissociate when the Debye screening of the color charge 
becomes so strong that it prevents the formation of bound states~\cite{Matsui:1986dk}.  
In recent years, the new suggestion is that the quarkonium states dissociate 
at a lower temperature~\cite{Laine:2006ns,Burnier:2007qm}, when the binding
energy of the quakonium is nonvanishing but exceeded by the thermal width~\cite{Laine:2006ns}, which is 
induced by the Landau damping and can be obtained~\cite{Laine:2006ns,Beraudo:2007ky,Laine:2007qy}
from the imaginary part of the $Q\bar Q$ potential.
The importance of the imaginary part of the potential has also been pointed out in recent perturbative calculations~\cite{Laine:2006ns,Burnier:2007qm}, namely, that the corresponding thermal width ($\Gamma$) of the $Q\bar Q$ bound states determines the dissociation temperature of the bound state.
The imaginary part of the potential has also been used earlier to construct the quarkonium spectral functions~\cite{Burnier:2007qm,Petreczky:2010tk} and calculate the perturbative thermal 
widths~\cite{Laine:2007qy, Brambilla:2010vq}.
Additionally, it has also been used to study the quarkonium properties using the $T$-matrix 
approach~\cite{Rapp:2008tf,Grandchamp:2005yw,Riek:2010py,Zhao:2011cv,Emerick:2011xu},
and using stochastic real-time dynamics~\cite{Akamatsu:2011se}.

In recent years there has been a renewed interest in the properties of bound 
states moving in a thermal medium due to the advent of high-energy heavy-ion colliders. The first study of the single heavy-quark potential and also 
the real part of the potential between a heavy quark and heavy antiquark pair 
moving with respect to the QGP was calculated in Ref.~\cite{Chu:1988wh} using a 
kinetic theory approach. Later, the screening potential of a single moving heavy quark was calculated by using semiclassical transport theory in Ref.~\cite{Mustafa:2004hf}. The formation of wakes created by a heavy moving parton in the QGP was studied~\cite{Abreu:2007kv,Ruppert:2005uz,Chakraborty:2006md,Chakraborty:2007ug,Jiang:2010zza}. The propagation of nonrelativistic bound states moving at constant velocity across a homogeneous thermal bath was studied in Ref.~\cite{Escobedo:2011ie}, in which the EFT was developed, which is relevant in various dynamical regimes. The authors of Ref.~\cite{Escobedo:2011ie} studied for the first time the imaginary part of the potential between a quark and antiquark moving in a thermal bath. Later, the decay width was also calculated from the imaginary part of 
the potential in Ref.~\cite{Escobedo:2013tca}. The authors of Ref.~\cite{Escobedo:2013tca} studied the modifications of some properties of weakly coupled heavy quarkonium states propagating through a QGP at temperatures much smaller than the heavy quark mass $m_{Q}$, and the two different hierarchies were considered. The first hierarchy corresponds to $m_Q \gg 1/r \gg T \gg E \gg m_D$, which is relevant for moderate temperatures, where $r$ is 
the size of the bound state, $E$ is the binding energy, $T$ is the temperature, and 
$m_D$ is the screening mass.
The second hierarchy corresponds to  $m_Q \gg T \gg 1/r,\ m_D \gg E$, which is relevant for studying the dissociation mechanism~\cite{Escobedo:2013tca}.
Recently, the thermal width of heavy quarkonia moving with respect to the
QGP was calculated up to the next-to-leading order (NLO) in perturbative QCD in Ref.~\cite{Song:2007gm}. At the leading order, the width decreases with increasing speed, whereas at the NLO it increases with a magnitude approximately proportional to the expectation value of the relative velocity
between the quarkonium and a parton in thermal equilibrium.
Additionally, we note that the heavy $Q\bar Q$ potential in a moving medium from holographic QCD can be found in Refs.~\cite{Liu:2006nn, Chernicoff:2006hi,Avramis:2006em,Caceres:2006ta,Liu:2012zw,Liu:2006he,Chakraborty:2012dt,Ali-Akbari:2014vpa,Patra:2015qoa,Noronha:2009da,Finazzo:2013aoa,Finazzo:2014rca}.
But whether AdS/CFT calculations represent real QCD phenomena remains controversial. Therefore we follow the weak coupling technique to calculate the complex potential and decay width of a quarkonium in a moving QGP medium.

Most of the analytic calculations that are available in the 
literature~\cite{Dumitru:2007hy,Dumitru:2009fy,Dumitru:2010id,Escobedo:2011ie,Escobedo:2013tca} using perturbative approaches to calculate the real and the imaginary parts of the heavy quark-antiquark potential were done only by considering the thermal modification of the vacuum Coulomb potential. Numerical lattice calculations have suggested that the transition from confined nuclear matter to deconfined matter is a crossover~\cite{Karsch:2006xs}, rather than a phase transition. So, the confinement term in the Cornell potential should be nonvanishing even above the transition temperature. Moreover, to study the large-distance properties (which helps us to understand the various bulk properties of
the QGP medium~\cite{Kaczmarek:2004gv,Dumitru:2004gd,Dumitru:2003hp}), the confinement potential term is necessary.   
So it is reasonable to study the effect of the string term above the deconfinement 
temperature~\cite{Cheng:2008bs,Maezawa:2007fc,Andreev:2006eh}.
Recently, the real~\cite{Agotiya:2008ie,Thakur:2012eb} and imaginary parts~\cite{Thakur:2013nia} of the potential have been calculated by modifying both the Coulombic and linear string terms of the Cornell potential in the static medium. The real and imaginary parts of the static interquark potential at finite temperature were derived in Ref.~\cite{Burnier:2015nsa}
in a weakly interacting medium by considering both perturbative and nonperturbative string terms.

Most of the calculations done so far to study heavy $Q\bar Q$ bound-state properties were done with a vanishing chemical potential of the thermal medium. But there have been some calculations with a nonvanishing chemical potential in both perturbative and non-perturbative approaches. For example, the color-screening
effect at both finite temperature and chemical potential was studied in thermo-field dynamics approach~\cite{Gao:1996xz,Liu:1997tc} to calculate the Debye screening mass, where the phenomenological potential model~\cite{Karsch:1987pv} and an error-function-type confined force with a color-screened  Coulomb-type potential was used. Lattice QCD has also been used to style the color screening in the heavy-quark potential at finite density with Wilson fermions~\cite{Takahashi:2013mja},
although it has severe limitations at finite chemical potential. Recently, properties of quarkonium states have also been studied perturbatively in a static medium at finite chemical potential~\cite{Kakade:2015laa} .

In the present work we study the behavior of the real and the imaginary parts of the potential between a heavy quark and its antiquark, which is in relative motion with respect to the thermal bath. In order to calculate the complex potential, we modify both the perturbative and nonperturbative terms of the vacuum potential through the dielectric function in the real-time formalism using the hard thermal loop (HTL) approximation. Each potential term (real or imaginary) depends on the angle between the orientation of the
dipole and the direction of motion of the thermal medium. In our calculation we consider two particular cases: when the heavy quarkonium state is aligned (i) along or (ii) perpendicular to the direction of the moving thermal medium. We also construct a general expression for the potential at any angle between the orientation of the
dipole and the direction of the relative motion of the thermal medium. 
It is important to discuss whether the dissociation mechanism (screening 
versus Landau damping) remains the same when the bound state moves
with respect to the thermal bath. Therefore, we calculate the decay width from the imaginary part of the potential in perturbation theory and also show the effect of the string term on it. We also study the chemical potential dependence of the various quantities that we are interested in.

This paper is structured as follows. In Sec.~\ref{sec:general_framework} we introduce the general framework for the properties of the moving particles in a deconfined thermal medium. In Sec.~\ref{sec:self_energy} we discuss the self-energies and propagators in a moving medium. In Sec.~\ref{sec:mass} we discuss the mass prescription of the Debye mass that we use here. In Sec.~\ref{sec:pot} we study the medium-modified heavy-quark potential in a moving thermal bath and also discuss the variation of the real and imaginary parts of the potential with various velocities when the $Q\bar{Q}$ pair is along/perpendicular to the direction of the velocity of the thermal medium. In Sec.~\ref{sec:gamma} we study the thermal decay width, and we present our conclusion in Sec.~\ref{sec:conclusion}.

\section{General Framework}
\label{sec:general_framework}
We consider a reference frame in which the QGP medium moves with a velocity ${\bf v}$ and the $Q\bar Q$ bound state is at rest. The particle distribution functions can be written 
as~\cite{Escobedo:2011ie}
\begin{equation}
f (\beta^\mu P_\mu ) =\frac{1}{e^{|\beta^\mu P_\mu|}\pm1},
\label{boosted-distributions}
\end{equation}
where the $+ (-)$ sign corresponds to fermions (bosons) and 
\begin{equation} 
\beta^\mu = \frac{\gamma}{T}(1,{\bf v})= \frac{u^\mu}{T} \,, 
\end{equation}
where $\gamma= 1/\sqrt{1-v^2}$ $(v=|{\bf v}|)$ is the Lorentz factor. 
We use the following notation for the four- and three- momenta in the above equation 
and also in the remaining part of the article: $P=(p_0,{\bf p}),\ p=|{\bf p}|$. 
The moving plasma frame was successfully used long ago in Ref.~\cite{Weldon:1982aq}.
Studying quakonium in a moving medium is equivalent to studying quakonium in nonequilibrium field theory~\cite{Carrington:1997sq}. In the nonequilibrium case the equilibrium Fermi-Dirac or Bose-Einstein distribution functions are  replaced by nonequilibrium distribution functions, whereas in the case of a moving thermal 
 medium the distribution functions will be the boosted Fermi-Dirac or Bose-Einstein distribution functions~(\ref{boosted-distributions}). In the nonequilibrium field theory we can write~\cite{Escobedo:2011ie}
\be
\label{dist}
\beta^\mu P_\mu=p\frac{1- v \cos{\theta}}{T\sqrt{1-v^2}}\,,
\ee
where $\theta$ is the angle between $\bf{p}$ and 
$\bf{v}$. The distribution functions 
(\ref{boosted-distributions}) can be written as 
\be
f(p,T, \theta, v)=\frac{1}{e^{p/T_{\rm eff}(\theta,v)}\pm 1}\,,
\label{boosted-2} 
\ee
where the {\it effective temperature} can be defined as
\be\label{effective-temperature}
T_{\rm eff}(\theta,v)=\frac{T\sqrt{1-v^2}}{1-v\cos{\theta}}\,.
\ee

The dependency on $v$ and $\theta$  in the effective temperature can be understand as a Doppler effect. For small values of $v$ ($v \ll 1$), $T_{\rm eff}(\theta,v) \sim T$ and the Boltzmann factor in Eq.~(\ref{boosted-2}) does not depend on the angle $\theta$; rather, it depends on only one scale, $T$. On the other hand, when $v$ is close to unity, the values of $T_{\rm eff}(\theta,v)$  depend on $\theta$. The number of scales introduced by moving the thermal medium can be understood by using light-cone coordinates~\cite{Escobedo:2011ie,Escobedo:2013tca}. In light-cone coordinates the distribution function depends on two scales, i.e., $T_{+}$ and $T_{-}$.  
Here we assume that both the scales $T_{+}$ and $T_{-}$  are of the same order of magnitude, as considered in Ref.~\cite{Escobedo:2013tca}.

\section{Self-Energies and Propagators}
\label{sec:self_energy}
To study the effect of the moving medium on the properties of the quarkonium states,
one first needs to calculate the self-energies and propagators in a moving medium.
We can calculate the self-energies and propagators in the real-time formalism~\cite{Carrington:1997sq} in the HTL approximation by assuming that the temperature of the plasma $ T \gg 1 /r  $. As we shall see in Sec.~\ref{sec:pot}, the heavy quark-antiquark potential is the Fourier transform of the physical ‘‘11’’ component of the gluon propagator in the real time formalism in the static limit. So, the ``11'' component of the longitudinal gluon propagator can be written in terms of the retarded 
($ D_R $), advanced ($ D_A $) and symmetric ($ D_S $) propagators in the real time formalism in the static limit as
 \begin{equation}
 D^{L}_{11}(p_0=0, {\bf p})=\frac{1}{2}\Big[D^{L}_R({\bf p})+D^{L}_A({\bf p})+D^{L}_S({\bf p})\Big] \,.
 \end{equation}

The retarded (advanced) propagator can be obtained from the retarded (advanced) self-energy and the symmetric propagator can be obtained from the symmetric self-energy. For the bound state  moving through the thermal bath, the symmetric 
propagator~\cite{Escobedo:2011ie} can be written as 
\be
\label{noeq}
D_S^L({\bf p},u)=\frac{\Pi_S^L({\bf p},u)}{2i\Im\Pi_R^L({\bf p},u)}\Big[D_R^L({\bf p},u) - D_A^L({\bf p},u)\Big] \,,
\ee
where $u^\mu=\gamma(1,{\bf v})$ is the four-velocity and $\Re$ and $\Im$ represent the real and imaginary parts, respectively. $D_R({\bf p},u)[D_A({\bf p},u)]$ represents the retarded [advanced] propagator for the bound state moving through the thermal bath. In order to determine the propagator, one has to evaluate the self-energies $\Pi_R({\bf p},u)$ and $\Pi_S({\bf p},u)$. The retarded self-energy $\Pi_R({\bf p},u)$ was computed 
in Ref.~\cite{Chu:1988wh} and later on in Ref.~\cite{Escobedo:2011ie} for the reference frame where the bound state is at rest, which can be written as 
 \begin{eqnarray}
\Pi_{R}({\bf p},u)=\Pi_{R}^{L}({\bf p},u)=a(z)+\frac{b(z)}{1-v^2}\, ,
\label{retself}
 \end{eqnarray} 
 where
 \begin{equation}
 a(z)=\frac{m_{D}^2}{2} \left[z^2-(z^2-1)\frac{z}{2}\log\(\frac{z+1+ i 
\epsilon}{z-1+ i \epsilon}\)\right],
 \label{a}
 \end{equation}
and
 \begin{eqnarray}
 b(z)=(z^2-1)\left[a(z)-m_{D}^2(1-z^2)\left\{1-\frac{z}{2}\log\(\frac{z+1+ i 
\epsilon}{z-1+ i \epsilon}\) \right\}\right],
 \label{b}
 \end{eqnarray}
 with
\be
 z=\left|\frac{P.u}{\sqrt{(P.u)^2-P^2}}\right|_{\omega=0}.
\label{z}
 \ee
 
 We always consider the dipole to be aligned along $z$ direction. If the velocity  direction is parallel to the axis of the dipole, Eq.~(\ref{z}) 
can be written as
 \be
 z_\parallel = \frac{v\cos\theta}{\sqrt{1-v^2\sin^2\theta}},
 \label{z_para}
 \ee
 where $(\theta,\phi)$ represent the polar and azimuthal angles of the momentum vector, respectively. Similarly, if the relative velocity of the medium is in the $x-y$ plane in a direction that makes an angle $\beta$ with the $x$ axis, Eq.~(\ref{z}) can be written as
  \be
 z_\perp = \frac{v\sin(\theta)\cos(\phi-\beta)}{\sqrt{1 - v^2 -v^2\sin^2\theta 
\cos^2(\phi-\beta)}}.
  \label{z_perp}
 \ee
 
Using Eqs.~(\ref{retself}) - (\ref{z_perp}), the retarded gluon
self-energy in the above-mentioned two directions can be written as
                             %
\begin{eqnarray}
\Pi^\parallel_{R}(\theta,v) &=&\frac{m_{D}^2}{2}\Bigg[ 
\frac{2-2v^2-v^4\cos^2\theta \sin^2\theta}{(1-v^2\sin^2\theta)^2}
-\frac{(2+v^2\sin^2\theta)(1-v^2)v\cos\theta}{2(1-v^2\sin^2\theta)^{5/2}}\nn
&& 
\hspace{4.cm}\times\log\left(\frac{v\cos\theta+\sqrt{1-v^2\sin^2\theta}}{
v\cos\theta-\sqrt{1-v^2\sin^2\theta}}
\right)\Bigg], 
 \end{eqnarray}
and
                                  %

\begin{eqnarray}
\Pi^\perp_{R}(\theta,\phi,\beta,v) &=&\frac{m_{D}^2}{2}\Bigg[ 
\frac{2-2v^2-v^4\cos^2(\phi-\beta)\sin^2\theta\(
1-\cos^2(\phi-\beta)\sin^2\theta\)}{(1-v^2 + 
v^2\cos^2(\phi-\beta)\sin^2\theta)^2}\nn
&&\hspace{1cm}-\frac{(2 +v^2 - 
v^2\cos^2(\phi-\beta)\sin^2\theta)(1-v^2)v\cos(\phi-\beta)\sin\theta}
{2(1-v^2 + v^2\cos^2(\phi-\beta)\sin^2\theta)^{5/2}}\nn
&&\hspace{1.5cm}\times\log\left(\frac{v\cos(\phi-\beta)\sin\theta
+ \sqrt{1-v^2 + v^2\cos^2(\phi-\beta)\sin^2\theta}}{v\cos(\phi-\beta)\sin\theta
- \sqrt{1-v^2 + v^2\cos^2(\phi-\beta)\sin^2\theta}}
\right)\Bigg],
 \end{eqnarray}
 where $m_D$ represents the Debye mass.
 
Using the retarded self-energy, one can obtain the retarded propagator as
\begin{equation}
D_{R}^{\para(\perp)}({\bf p},u)=D_{R}^{L,\ \para(\perp)}({\bf p},u)\frac{1}{p^2+\Pi_R^{\para(\perp)}({\bf p},u)}
.
\end{equation}

For a bound state in the moving thermal bath it is enough to calculate the 
retarded self-energy and the advanced self-energy can be obtained from the relation 
$D_{R}^{*\para(\perp)}({\bf p},u)=D_{A}^{\para(\perp)}({\bf p},u)  $. 
Similarly, one can calculate the symmetric self-energies for both cases as 
\begin{eqnarray}
 \Pi_S^\para({\bf p},u)=\Pi^{L,\para}_{S}=\Pi_1+\frac{\Pi_2}{1-v^2}
 =\frac{i2\pi m_D^2 T(1-v^2)^{3/2}\(1+\frac{v^2}{2}\sin^2\theta\)}{|{\bf 
p}|\(1-v^2\sin^2\theta\)^{5/2}}\,
 \label{Pis00_pr}
 \end{eqnarray}
 and
  \begin{eqnarray}
 \Pi_S^\perp({\bf p},u)=\Pi^{L,\perp}_{S}=\Pi_1+\frac{\Pi_2}{1-v^2}
 =\frac{i 2\pi m_D^2 
T(1-v^2)^{3/2}\(1+\frac{v^2}{2}-\frac{v^2}{2}\cos^2(\phi-\beta)\sin^2\theta\)}{|{
\bf p}|\(1 - v^2 
 + v^2 \cos^2(\phi-\beta)\sin^2\theta\)^{5/2}}\,.
 \label{Pis00_pp}
 \end{eqnarray}
 
The symmetric propagator $ D_S^{L}({\bf p},u) $ can be obtained by substituting the symmetric self-energy into Eq.~(\ref{noeq}) as
 \begin{eqnarray}
D_S^{\para(\perp)}({\bf p},u)=\frac{\Pi_S^{\para(\perp)}({\bf p},u)}{2i\Im\Pi_R^{
\para(\perp)}({\bf p},u)}\left( 
D_{R}^{\para(\perp)}({\bf p},u)-D_{A}^{\para(\perp)}({\bf p},u)\right) \,,
 \end{eqnarray}
 
 Because $D_{A}({\bf p},u)= D_{R}^*({\bf p},u)$, $D_{R}({\bf p},u)-D_{A}({\bf p},u)$ can be written as
\be
 D_{R}^{\para(\perp)}({\bf p},u)-D_{A}^{\para(\perp)}({\bf p},u)&=&\frac{1}{p^2 + 
\Pi_R^{\para(\perp)}({\bf p},u)} - 
 \frac{1}{p^2 + \Pi_R^{*\para(\perp)}({\bf p},u)}\nn
 &&= \frac{2i\Im\Pi_R^{\para(\perp)}(P,u)}{\(p^2 + \Pi_R^{\para(\perp)}({\bf p},u)\)
 \(p^2 + \Pi_R^{*{\para(\perp)}}({\bf p},u)\)}.
\ee
Therefore, we get the symmetric propagator for the parallel case as
\begin{eqnarray}
D_S^\para({\bf p},u)=D_S^{L,\para}({\bf p},u)=\frac{-2\pi i m_D^2 T(1-v^2)^{3/2}(2+v^2\sin^2\theta)}{2p(1-v^2\sin^2\theta)^{5/2}
\(p^2 + \Pi_R^\para({\bf p},u)\)\(p^2 + \Pi_R^{*\para}({\bf p},u)\)} \,
 \end{eqnarray}
 and for the perpendicular case as
\begin{eqnarray}
 D_S^\perp({\bf p},u)=D_S^{L,\perp}({\bf p},u=\frac{-2\pi i m_D^2 T(1-v^2)^{3/2}\(2+v^2-v^2\cos^2(\phi-\beta)\sin^2\theta\)}
 {2p\(1-v^2+v^2\cos^2(\phi-\beta)\sin^2\theta)\)^{5/2}\(p^2 + \Pi_R^\perp({\bf p},u)\)\(p^2 + \Pi_R^{*\perp}({\bf p},u)\)} \ .
 \nn
\end{eqnarray}

Using these above expressions for the self-energies and propagators, in Sec.~\ref{sec:pot} we calculate the heavy-quark potential in a moving medium.
 
\section{Mass prescription}
\label{sec:mass}
We use the NLO Debye mass as given in Ref.~\cite{Haque:2014rua} and 
the expression for $m_D^2$ can be written at finite chemical potential ($ \mu\neq 
0 $) as
\begin{eqnarray}
 m_D^2&=&\frac{4\pi \alpha_s}{3}T^2\Bigg[N_c +\frac{N_c^2\alpha_s}{12\pi}\(5+22\gamma_E
 + 22\lL\) + \frac{N_f}{2}\(1+12\hmu^2\) \nn
&& +\frac{N_cN_f\alpha_s}{24\pi} \Bigg\{\(9+132 \hmu^2 \) +22 \(1+12 \hmu^2 
\)\gamma_E  + \, 2\( 7 + 132\hmu^2\)\lL+4\aleph(\hmu)\Bigg\}\nn
&&+\frac{N_f^2\alpha_s}{12\pi}\(1+12\hmu^2\)\(1-2\lL+\aleph(\hmu)\)-\frac{3}{8\pi}
\frac{N_f(N_c^2-1)\alpha_s}{N_c}\(1+12\hmu^2\)\Bigg] ,
\label{eq:bnmass}
\end{eqnarray}

where $N_c$ represents the number of colors, $N_f$ represents the number of flavors, $\hmu=\mu/(2\pi T)$, $\aleph(\hmu)=\psi(1/2-i\hmu) + \psi(1/2+i\hmu)$, and $\psi$ denotes digamma function. The expression for the running coupling $\alpha_s$ at the one loop level can be written as
\be
\alpha_{s}&=&\frac{12\pi}{11N_c - 2N_f}\frac{1}{ 
\ln(\Lambda^2/\Lambda_{\overline{\rm MS}}^2)}.
\ee
We use $\Lambda_{\overline{\rm MS}} = 176$ MeV and the renormalization scale $\Lambda=2\pi\sqrt{T^2+\mu^2/\pi^2}$, as discussed in Ref.~\cite{Haque:2014rua}.

The quantities that we discuss in this article depend on the quark chemical potential only via the Debye mass. In most of the numerical plots we take a vanishing chemical potential, but in some cases we also take a finite chemical potential. Whenever there is a finite value for the chemical potential, we mention it. If the value of $\mu$ is not mentioned, a vanishing chemical potential is implied.
 
 \section{Heavy Quark-antiquark potential in a moving medium}
 \label{sec:pot}
In this section we study the modification of the full Cornell potential of a heavy quark-antiquark which is in relative motion with respect to the thermal bath. The modification of the purely Coulombic potential of a moving heavy quark-antiquark has been studied in Refs.~\cite{Escobedo:2011ie,Escobedo:2013tca}. We evaluate the potential in the HTL approximation by assuming that the temperature of the plasma $T\gg 1/r$.
The medium-modification to the vacuum potential can be obtained
by correcting both the perturbative (short-distance) and nonperturbative (long-distance) parts with a dielectric function $ \epsilon(p)$ encoding the effect of deconfinement~\cite{Agotiya:2008ie,Thakur:2012eb},
\begin{eqnarray}
\label{defn}
V({\bf r},T,{\bf v})&=& \int \frac{d^3\mathbf p}{{(2\pi)}^{3/2}}
 (e^{i\mathbf{p} \cdot \mathbf{r}}-1)~\frac{V(p)}{\epsilon(p,u)} ~,
 \end{eqnarray}
 where $r$-independent term (perturbative free energy term of a quarkonium at infinite separation~\cite{Dumitru:2009fy}) has been subtracted to renormalize the heavy quark free energy. $ V(p) $ is the Fourier transforms (FT) of the Cornell potential which can be written from Ref.~\cite{Agotiya:2008ie} as
 \be
 \label{vp}
 {V}(p) &=&-\sqrt{(2/\pi)} \frac{\alpha}{p^2}-\frac{4\sigma}{\sqrt{2 \pi} p^4}\nn
 &=&-\sqrt{(2/\pi)} \frac{C_F\alpha_s}{p^2}-\frac{4\sigma}{\sqrt{2 \pi} p^4},
 \ee
 where $ \alpha \equiv C_F \alpha_s$ [with $ C_F=(N_c^2-1)/2N_c $] and $\sigma$ represents the string tension. The dielectric permittivity  $ \epsilon (\mathbf{p},u) $ 
can be obtained perturbatively from the relation~\cite{Kapusta:2006pm}
 \begin{equation}
 \epsilon^{-1}({\bf p},u)=\lim_{\omega \to 0} {p^2} D^{L}_{11}(\omega, {\bf p},u),
 \label{ephs}
 \end{equation}
where $ D^{L}_{11} $ is the longitudinal gluon propagator. Note that we are calculating the dielectric permittivity perturbatively, though $V(p)$ in Eq.~(\ref{defn}) contains a nonperturbative linear term. So, at large distances or in the soft momenta region (where the potential is dominated by the nonperturbative term) this perturbative
approximation may not be appropriate for calculating the dielectric permittivity.

Now the longitudinal gluon propagator can be written in terms of the real and imaginary parts as
 \begin{eqnarray}
 D^{L}_{11}({\bf p},u)= \Re D^{L}_{L}({\bf p},u) + \Im D^{L}_{11}({\bf p},u),
 \label{F1}
 \end{eqnarray}
 The real part of the propagator can be written in terms of the retarded and advanced propagators, and the imaginary part can be written in terms of the symmetric propagator as
 \begin{eqnarray}
 \Re D_{11}^{L}({\bf p},u)= \frac{1}{2}\left( D^{L}_{R}({\bf p},u) + D^{L}_{A}({\bf p},u)\right)
{\rm{and}}~~
 \Im D_{11}^{L}({\bf p},u)= \frac{1}{2} D^{L}_{S}({\bf p},u).
 \label{D11}
 \end{eqnarray}
 Thus, by using the real and imaginary parts of the propagator we can calculate the real 
and imaginary parts of the heavy-quark potential as
 \begin{eqnarray}
V({\bf r},T,{\bf v})&=& \int \frac{d^3\mathbf p}{{(2\pi)}^{3/2}}(e^{i\mathbf{p} \cdot 
\mathbf{r}}-1)~V(p)\ p^2D_{11}^{L}({\bf p},u)\ \nn
&=&  \int \frac{d^3\mathbf p}{{(2\pi)}^{3/2}}(e^{i\mathbf{p} \cdot 
\mathbf{r}}-1)~V(p)\ \frac{p^2}{2}
   \left(D_R^{L}({\bf p},u)+D_A^{L}({\bf p},u)+D_S^{L}({\bf p},u))\right).
   \label{30}
 \end{eqnarray}
Note that we are using same screening scale for both the Coulombic and string terms, contrary to Ref.~\cite{Burnier:2015nsa} in which the authors used different screening scales. Now, Eq. (\ref{30}) can be decomposed into the real and imaginary part of the potential as
\be
V({\bf r},T,{\bf v})&=&  \Re V({\bf r},T,{\bf v})+\Im V({\bf r},T,{\bf v}),
\ee
where the real part is
\be
 \Re V({\bf r},T,{\bf v})&=&  \int \frac{d^3\mathbf p}{{(2\pi)}^{3/2}}\ \(e^{i\mathbf{p 
\cdot r}} -1 \)~V(p)\ \frac{p^2}{2}\(D_R^{L}({\bf p},u)
 +D_A^{L}({\bf p},u)\),
 \label{pot_re_def}
\ee
and the imaginary part is 
\be
 \Im V({\bf r},T,{\bf v})&=&  \int \frac{d^3\mathbf p}{{(2\pi)}^{3/2}}(e^{i\mathbf{p 
\cdot r}}-1)~V(p)\ \frac{p^2}{2}D_S^{L}({\bf p},u)\ .
 \label{pot_im_def}
\ee

Each potential term (real or imaginary) depends on the angle between the 
orientation of the dipole and the direction
of motion of the thermal medium.
 For simplicity, we consider two extreme cases:  
\begin{itemize}
 \item The dipole is aligned parallel to the direction of the relative velocity 
of the dipole and the medium.

 \item The dipole is aligned in the perpendicular plane of the direction of the 
relative velocity of the dipole and the medium.
\end{itemize}

Using the expressions for the potential in the parallel and perpendicular alignments, it is possible to construct a general expression for the potential at any angle between the orientation of the dipole and the direction of relative motion of the dipole and the medium. We give below the real and imaginary parts of the potential for both cases.

\subsection{Real part of the potential}
The real part of the potential can thus be obtained from Eq.~(\ref{pot_re_def}) as
 \begin{eqnarray}
 \Re V({\bf r},T,{\bf v})&=& \int \frac{d^3\mathbf p}{{(2\pi)}^{3/2}}\ \(e^{i\mathbf{p 
\cdot r}} - 1\)~V(p)\ \frac{p^2}{2}\(D_R^{L}({\bf p},u) 
 + D_A^{L}({\bf p},u)\)\nn
 &=&- \int \frac{d^3\mathbf p}{{(2\pi)}^{3/2}}\ \(e^{i\mathbf{p} \cdot 
\mathbf{r}}  -1 \)\left(\sqrt{(2/\pi)} C_F \alpha_s
   +\frac{4\sigma}{\sqrt{2 \pi} p^2}\right) 
\Re\left[\frac{1}{p^2 + \Pi_R({\bf p},u)}\right]
\end{eqnarray}
When the dipole is aligned parallel to the direction of the relative velocity, the gluon self-energy depends only on the polar angle $\theta$ due to the azimuthal symmetry. So, the real part of the potential in this alignment can be written as
\begin{eqnarray}
\Re V(\mathbf{r \parallel v} ,T) &= & - \frac{\alpha_s C_F }{r}\nn
&-&\alpha_s m_D C_F\ \Re\Bigg[\int\limits_0^{\pi/2} d\theta \sin\theta\ \(1 - e^{-m_D r \cos\theta 
\sqrt{\Pi^\para_R(\theta,v)/m_D^2}}\)\sqrt{\Pi^\para_R(\theta,v)/m_D^2}\Bigg]
\nn
 &+&\frac{\sigma}{2m_D}\ \Re\Bigg[ \int\limits_0^{\pi/2} d\theta \sin\theta\  
 \(1-e^{-m_D r \cos\theta \sqrt{\Pi^\para_R(\theta,v)/m_D^2}}\)\frac{1}{\sqrt{\Pi^\para_R
 (\theta,v)/m_D^2}}\Bigg].
 \label{re_pot_para}
\end{eqnarray}
    
\begin{figure}[tbh]
\subfigure{
\hspace{-0mm}\includegraphics[width=7.5cm]{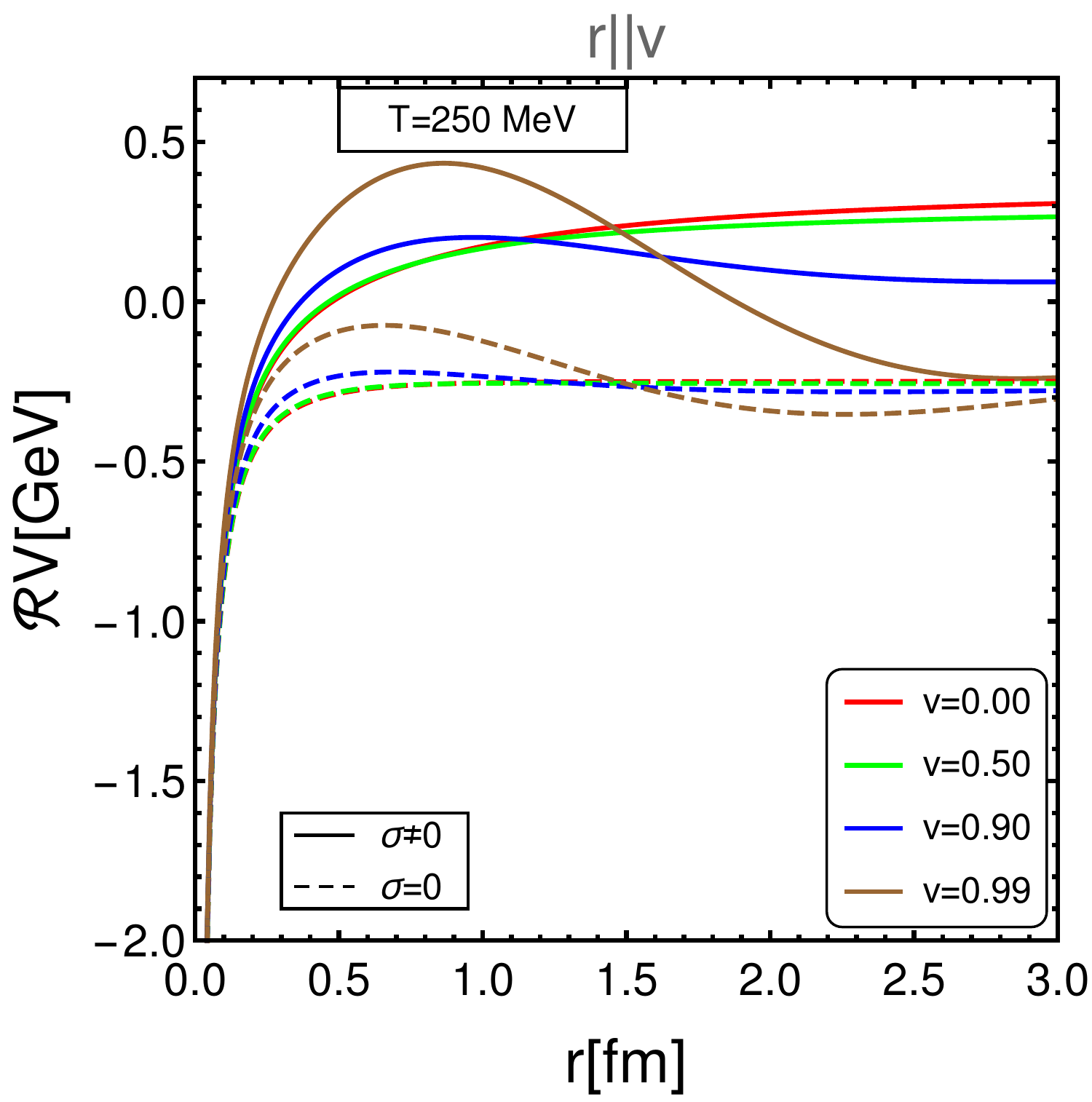}} 
\subfigure{
\includegraphics[width=7.5cm]{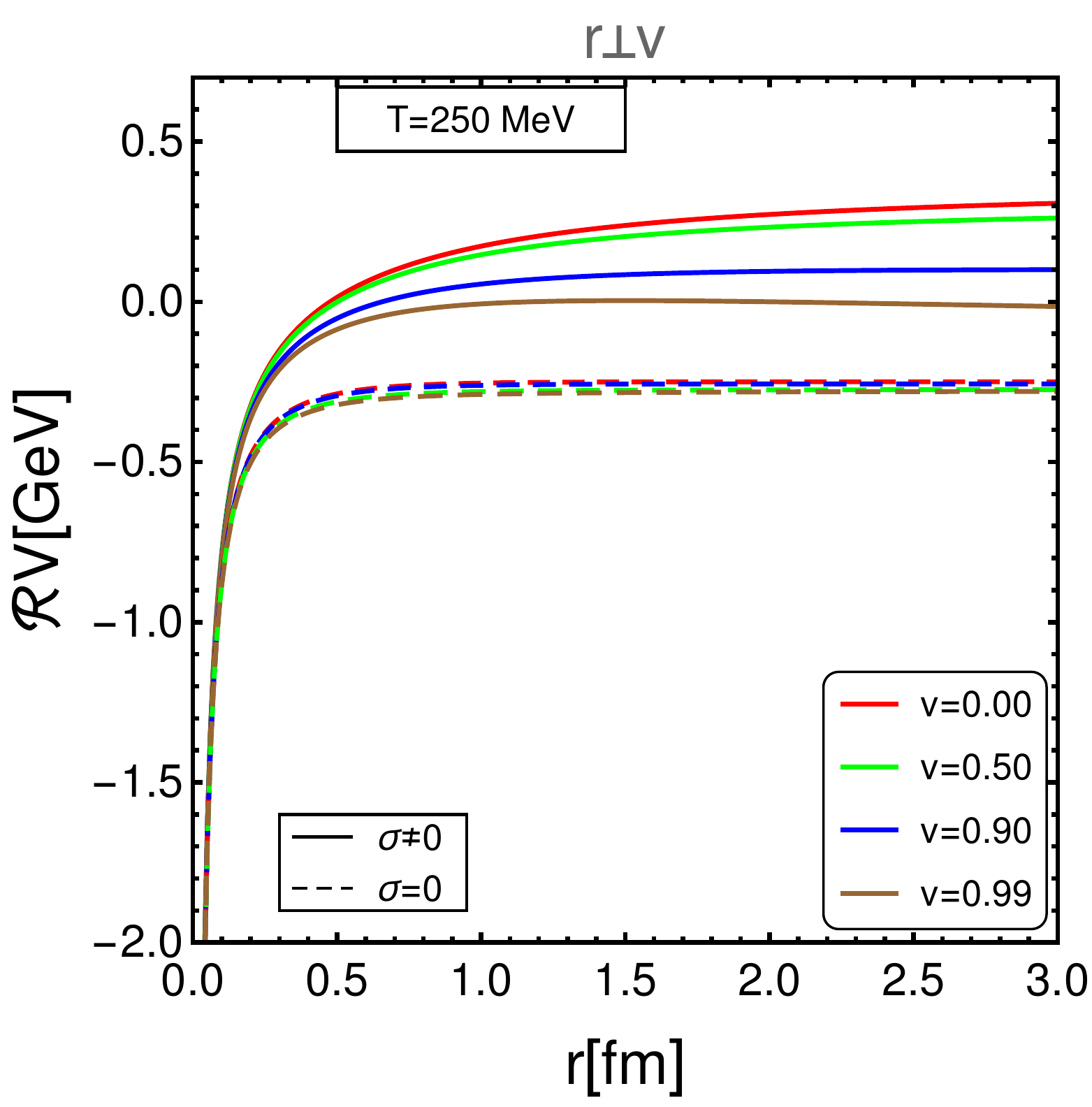}}
\caption{
Variation of the real part of the potential with the separation distance $ r $ between $Q$ 
and $\bar Q$ for various values of velocity at $T=250$ MeV, when the dipole axis
is along (left) and perpendicular to (right) the direction of the velocity of the thermal medium.}
\label{fig:pot_re_rmD}
\end{figure}
\noindent 
But, in the perpendicular alignment, the gluon self-energy depends on the polar angle $\theta$, the azimuthal angle $\phi$, and the angle between the velocity and $x$ axis: the real part of the potential can be written as
\begin{eqnarray}
\Re V(\mathbf{r\! \perp\! v} ,T)&=&  -\frac{\alpha_s m_D C_F}{ r} \nn
&-&\alpha_s m_D C_F\Re\Bigg[ \int\limits_0^{\pi/2}\!\! d\theta \sin\theta\! 
\int\limits_0^{2\pi}\frac{d\phi}{2\pi}\(1 - e^{-m_D r \cos\theta 
\sqrt{\Pi^\perp_R(\theta,\phi,\beta,v)/m_D^2}}\)
 \sqrt{\Pi^\perp_R(\theta,\phi,\beta,v)/m_D^2}\Bigg]\nn
 & + &\frac{\sigma}{2m_D} \Re\Bigg[ \int\limits_0^{\pi/2}\! d\theta \sin\theta 
\int\limits_0^{2\pi}\frac{d\phi}{2\pi} \(1 -  
e^{-m_D r \cos\theta \sqrt{\Pi^\perp_R(\theta,\phi,\beta,v)/m_D^2}}\)\frac{1}
 {\sqrt{\Pi^\perp_R(\theta,\phi,\beta,v)/m_D^2}}\Bigg].\ \nn 
 \label{re_pot_perp}
  \end{eqnarray}
\begin{figure}[tbh]
\subfigure{
\hspace{-0mm}\includegraphics[width=7.5cm]{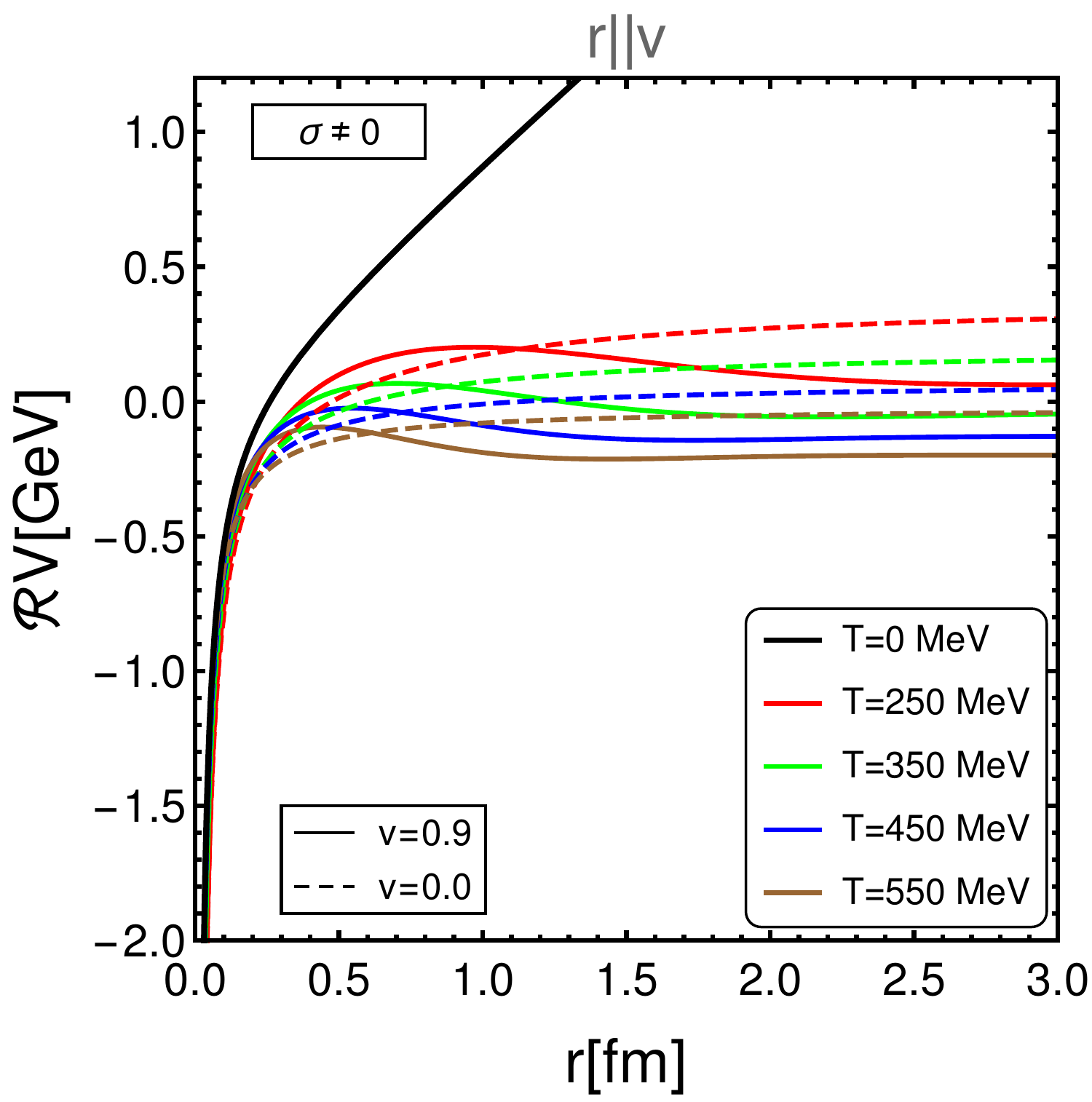}} 
\subfigure{
\includegraphics[width=7.5cm]{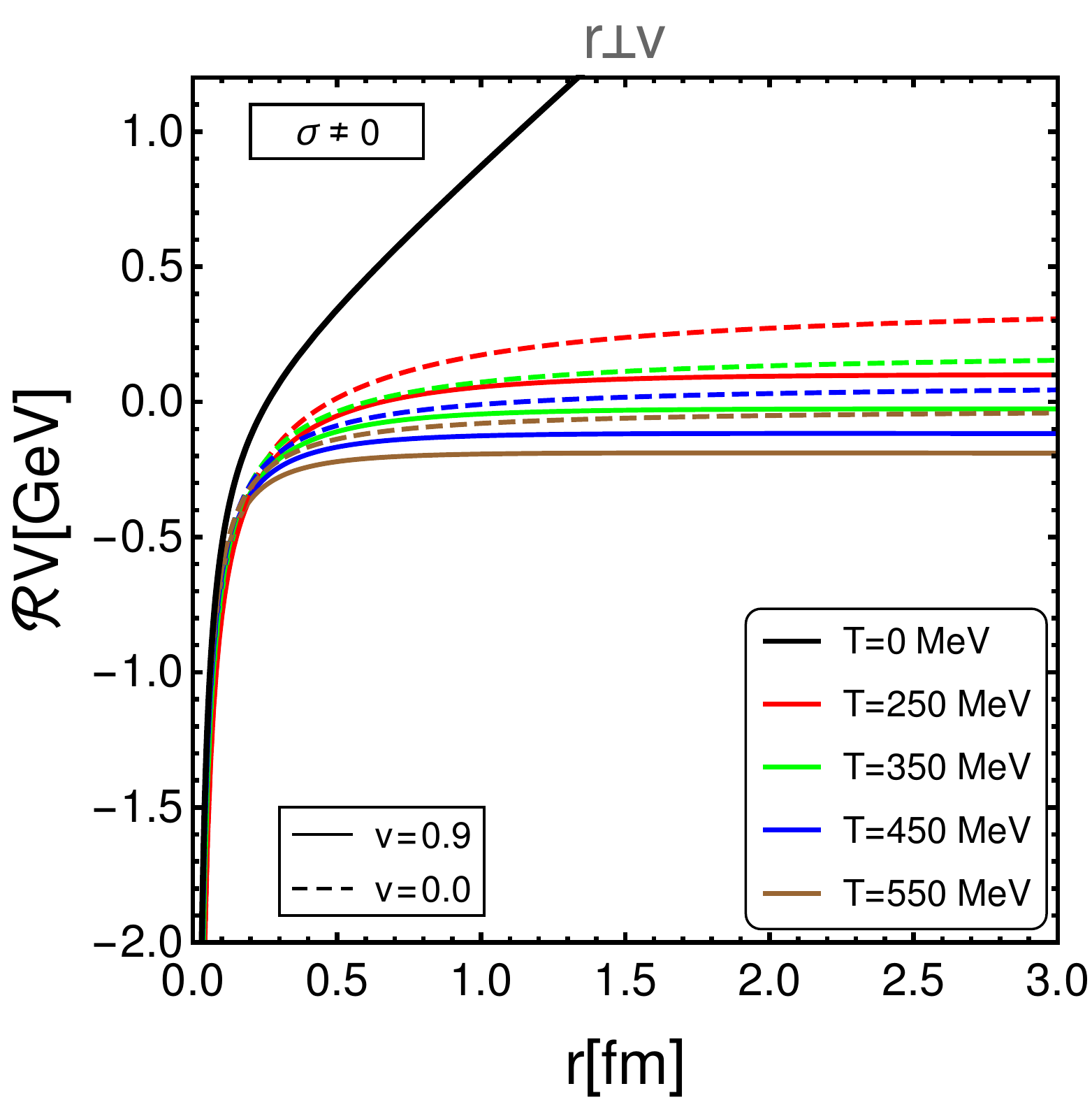}}
\caption{
Variation of the real part of the potential with the separation distance $ r $ between $Q$ 
and $\bar Q$ for various values of temperature at  $ v=0 $ (dotted line) and  $ v=0.9 $ (solid line),  when the dipole axis is along (left) and perpendicular to (right) the direction of the velocity of the thermal medium, along with the $T=0$ (solid black line) potential.}
\label{fig:pot_re_v}
\end{figure}
Figure~\ref{fig:pot_re_rmD} shows the variation of the real part of the potential for various values of velocity ( $v=0.0,\,0.50,\,0.90,\, 0.99)$ at $T=250$ MeV for both 
orientations of the dipole. We use the value of the string tension $\sigma=0.184$ GeV$^2$ 
in this plot and the rest of the paper. From the figure we find that the real part of the potential increases with an increase in velocity at short distances, and decreases
with an increase in velocity at large distances. The increase in velocity makes the real part of the potential less screened at short distances and more screened at large distances for the parallel case. On the other hand, the potential decreases with an increase in velocity for the perpendicular case which results in more screening of the potential. The inclusion of the string term makes the real part of the potential less screened as compared to the Coulombic term alone for both cases. Combining all these effects, we expect a stronger binding of a $ Q\bar{Q} $ pair in the presence of the string term as compared to the Coulombic term alone in the moving medium. Here, solid lines represent the potential with both the Coulomb and string terms, whereas dashed lines represent the potential in the absence of the nonperturbative (string) term.
The real part of the potential shows an oscillation in the ultrarelativistic limit for the parallel case which may be because of the induced dipole interaction in this direction. 

Figure~\ref{fig:pot_re_v} shows the variation of the real part of the potential with the separation distance $r$ for different values of temperature ( $T=250,\, 350,\, 450,\,$ and $550\, \text{MeV})$ for both the parallel and perpendicular cases. In this figure, solid black lines represent the $T=0$ potential. Other solid lines represent the potential at $ v=0.9 $ and dashed lines represent the potential at $ v=0 $. From the figure we find that the real part of the potential becomes more screened with an increase in temperature for both  cases and becomes short range. Alternatively, we can say that at higher temperatures the $Q\bar Q$ state is loosely bound as compared to the lower temperatures.
The zero-velocity plots in the above two figures look quantitatively similar
to the other finite-temperature heavy quark-antiquark potentials available in 
the literature~\cite{Kaczmarek:2002mc,Kaczmarek:2004gv,Mocsy:2007yj,Dumitru:2009ni}.
Note that the potential does not depend on the angle $\beta$. This means that the potential is independent of the angle of orientation of the dipole in the $x-y$ plane. In both Figs.~\ref{fig:pot_re_rmD} and~\ref{fig:pot_re_v}, we take a vanishing chemical potential of the medium.

\begin{figure}[tbh]
\subfigure{
\hspace{-0mm}\includegraphics[width=7.5cm]{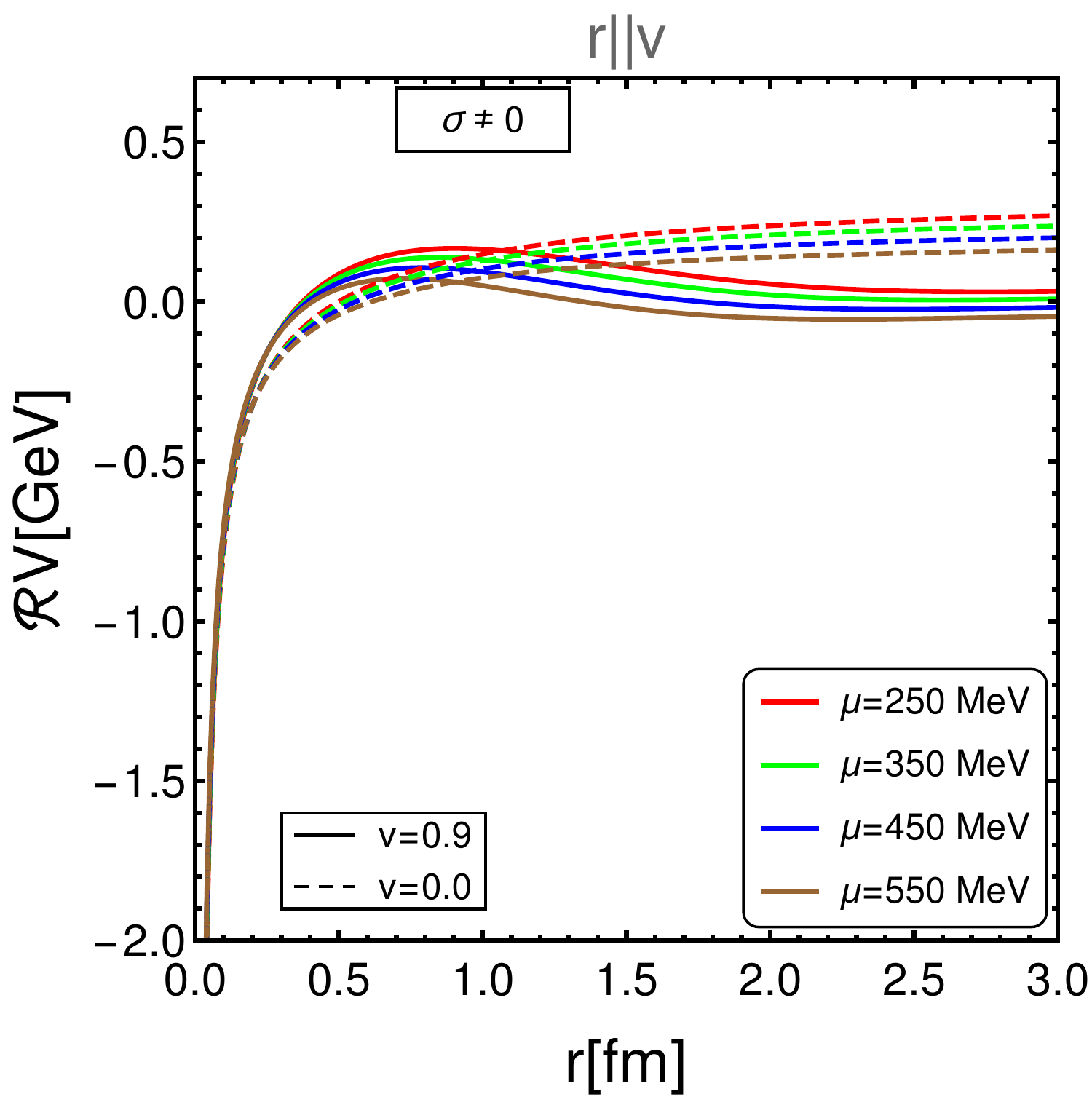}} 
\subfigure{
\includegraphics[width=7.5cm]{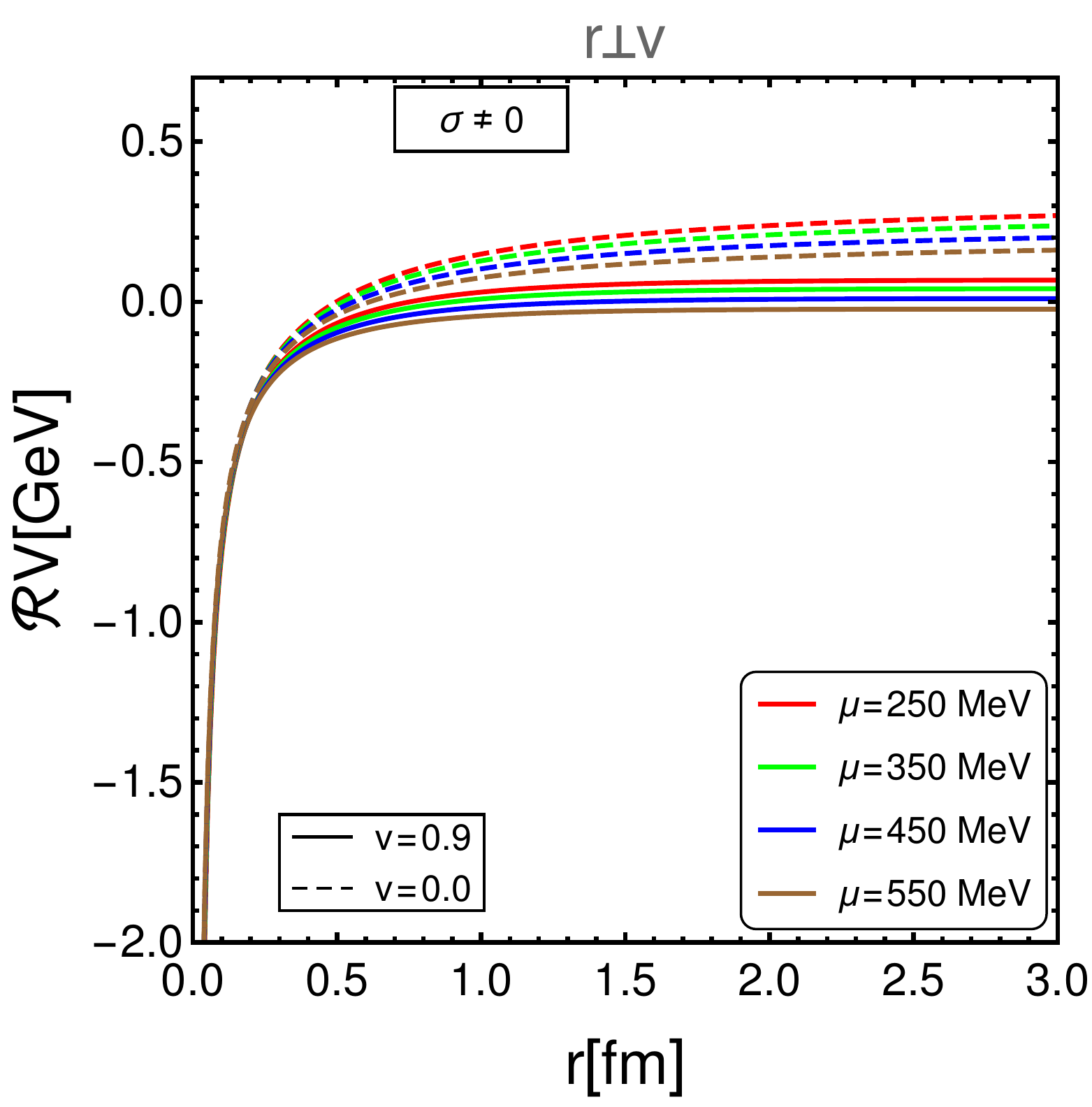}}
\caption{
Variation of the real part of the potential with the separation distance $ r $ between $Q$ 
and $\bar Q$ for various values of chemical potential at $ v=0 $ (dotted line) and  $ v=0.9 $ (solid line),  when the dipole axis is along (left) and perpendicular to (right) the direction of the velocity of the thermal medium.}
\label{fig:pot_re_vmu}
\end{figure}

Figure~\ref{fig:pot_re_vmu} shows the variation of the real part of the potential with the separation distance $ r $ for different values of chemical potential ( $\mu=250,\, 350,\, 450,\,$ and $ 550 \, \text{MeV})$ for both the parallel and perpendicular orientations. We find almost the same behavior of the potential with an increase in $ \mu $ as in 
Fig.~\ref{fig:pot_re_v} with temperature, but the potential shows a very weak dependence on $\mu $ as compared to the temperature $T$. This can be easily understood from the fact that the energy scale induced by the temperature
is $2\pi T$, while the one induced by the chemical potential is just $\mu$.

If the angle between the dipole axis and the velocity vector is $\theta_{vr}$, then the real 
part of the potential at any orientation of
the dipole can be written as
\be
\Re V(\mathbf{r } ,T,\mathbf{v},\theta_{vr})&=&  A(\mathbf{r } ,T,\mathbf{v}) + B(\mathbf{r } ,T,\mathbf{v}) \cos2\theta_{vr}
\label{eq:pot_re_theta_AB}
\ee
Here, we follow the same procedure that has been used to write the general angle dependence potential in the static
anisotropy medium at small anisotropy parameter in Refs.~\cite{Dumitru:2009ni,Thakur:2012eb}. So, 
Eq.~(\ref{eq:pot_re_theta_AB})
can only be trusted at small velocity. 
Now, one can extract the functions $A$ and $B$ from Eq.~(\ref{eq:pot_re_theta_AB}) by choosing $\theta_{vr}=0$ and
$\theta_{vr}=\pi/2$, and the real part of the potential at any orientation of
the dipole can be written as
\be
\Re V(\mathbf{r } ,T,\mathbf{v})&=& \frac{1}{2} \Re V(\mathbf{r \para v} ,T)\(1 
+\cos2\theta_{vr}\) +  \frac{1}{2} \Re V(\mathbf{r \perp v} ,T)\(1 - 
\cos2\theta_{vr}\). 
\label{eq:pot_re_theta}
\ee
\begin{figure}[tbh]
\subfigure{
\hspace{-0mm}\includegraphics[width=7.5cm]{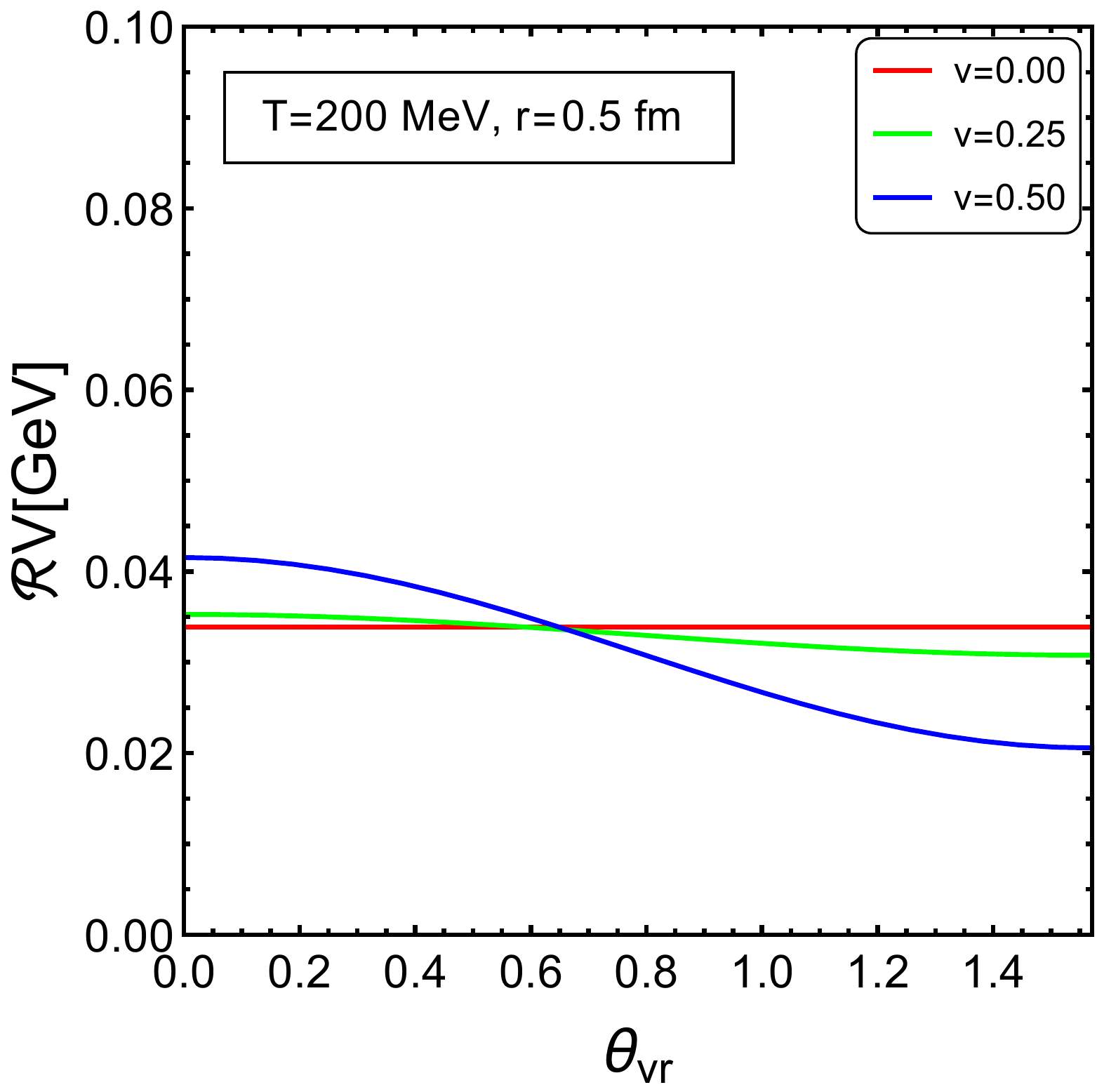}} 
\subfigure{
\includegraphics[width=7.5cm]{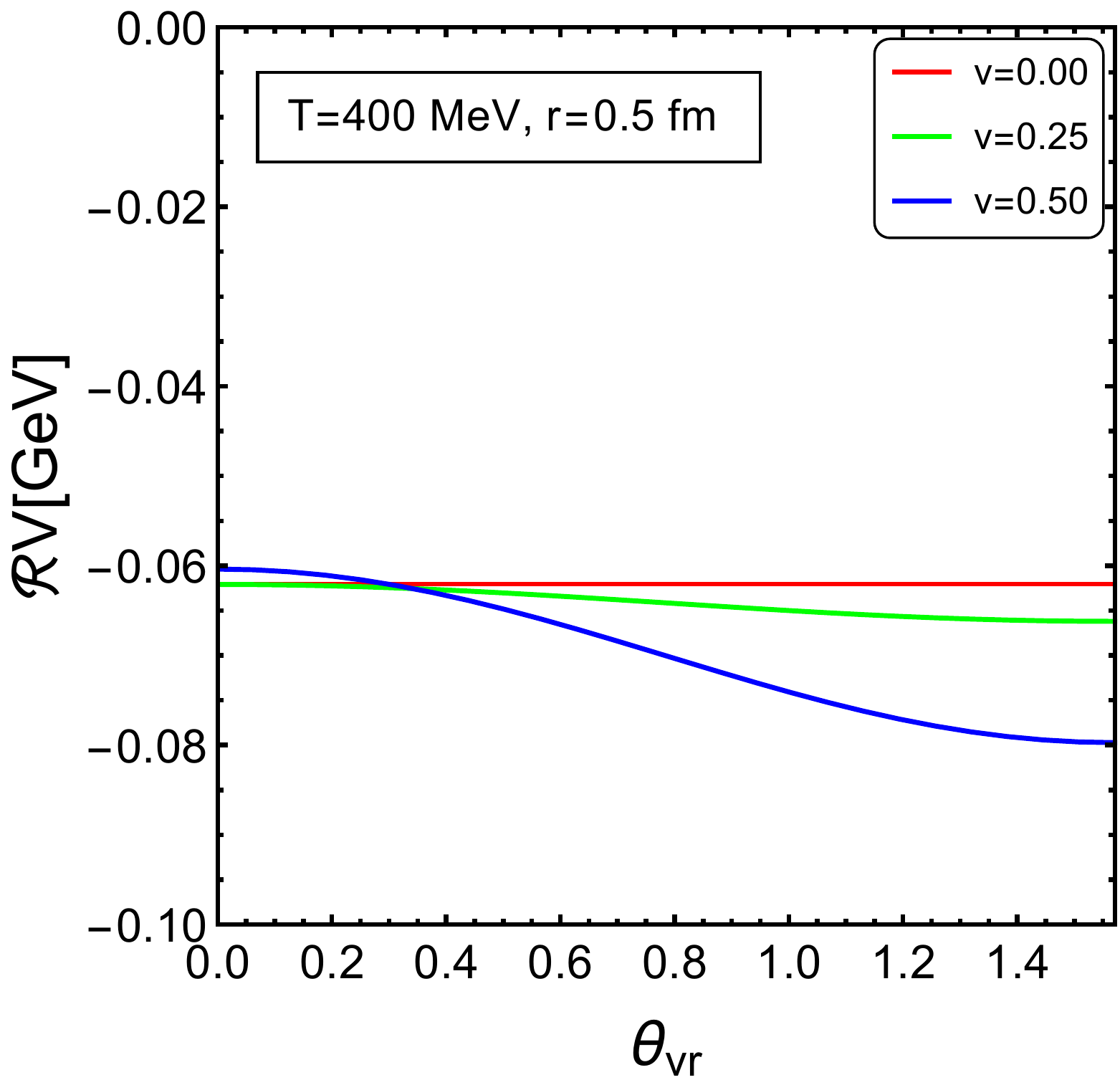}}
\caption{
Real part of the potential for various values of velocity at $T=200$ MeV (left) 
and $T=400$ MeV  (right) with the 
angle $\theta_{vr}$ between the dipole orientation and the velocity direction.}
\label{pot_re_theta}
\end{figure}

Figure~\ref{pot_re_theta} shows the variation of the real part of the potential with the
angle $ \theta_{vr} $ between the dipole orientation and the velocity direction.
We find that the real potential remains almost constant with $ \theta_{vr} $ at small velocities. It increases at moderate velocities near the parallel alignment and decreases with an increase in $ \theta_{vr} $, which due to the smaller screening near the parallel alignment as compared to near the perpendicular alignment. The increase in the value of the potential at moderate velocities is less at high temperature ($ T=400 $ MeV) as compared to low temperature ($ T=200 $ MeV), which is due to the larger screening at high temperature.

When the relative velocity between the $Q\bar Q$ pair and the thermal medium is 
small $(v\lesssim 0.5)$, it is possible to compute the real
part of the potential in the parallel and perpendicular directions from 
Eq.~(\ref{re_pot_para}) and Eq.~(\ref{re_pot_perp}), respectively, in terms 
of the scaled separation $\hat r(= r m_D)$ as
\be
\Re V(\mathbf{r \parallel v} ,T) &\approx & -\alpha_s m_D C_F 
\Bigg[1 + \frac{e^{-\hr}}{\hr}+ 
    v^2\Bigg\{-\frac{6}{\hr^3} + e^{-\hr} 
\left(\frac{6}{\hr^3}+\frac{6}{\hr^2}+\frac{3}{\hr}+1\right)
   \nn
&+& \ \frac{\pi^2}{2}\(\frac{1}{\hr^3}+\frac{1}{48}\) -\frac{e^{-\hr}\pi^2}{2}\left(\frac{1}{\hr^3}+\frac{1}
     {\hr^2}+\frac{1}{2\hr}+\frac{3}{16} +\frac{\hr}{16} \right)\Bigg\}\Bigg]\nn
&+& \frac{2\sigma }{m_D} \Bigg[1 + \frac{e^{-\hr}-1}{\hr}+ 
   v^2\Bigg\{-\frac{12}{\hr^3} + \frac{1}{\hr}+e^{-\hr} 
     \left(\frac{12}{\hr^3}+\frac{12}{\hr^2}+\frac{5}{\hr}+1\right)
     \nn
&+& \ \frac{\pi^2}{2}\(\frac{3}{\hr^3} - \frac{1}{16}\) -\frac{e^{-\hr}\pi^2}{2} 
\left(\frac{3}{\hr^3}+\frac{3}{\hr^2}+\frac{3}{2 \hr}+\frac{7}{16} +\frac{\hr}{16}
\right)\Bigg\}\Bigg] + \mathcal{O}(v)^4,
\label{pot_re_para_small_v}
   \ee
and
\be
\Re V(\mathbf{r \perp v} ,T) &\approx &  -\alpha_s m_D C_F \Bigg[ 1 + 
\frac{e^{-\hr}}{\hr}+ v^2\Bigg\{\frac{3}{\hr^3} - \frac{e^{-\hr}}{2} 
\left(\frac{6}{\hr^3}+\frac{6}{\hr^2}+\frac{3}{\hr}+1\right)
 \nn
&-& \ \frac{\pi^2}{4}\(\frac{1}{\hr^3} -\frac{1}{24}\) + \frac{e^{-\hr}\pi^2}{4} 
\left(\frac{1}{\hr^3}+\frac{1}{\hr^2}+\frac{1}{2\hr}+\frac{1}{8} \right)\Bigg\}\Bigg]\nn
&+& \frac{2\sigma}{m_D} \Bigg[ 1+ \frac{e^{-\hr}-1}{\hr}+ 
v^2\Bigg\{\frac{6}{\hr^3} - \frac{1}{2\hr} -\frac{e^{-\hr}}{2} 
     \left(\frac{12}{\hr^3}+\frac{12}{\hr^2}+\frac{5}{\hr}+1\right)
    \nn
&+&  \ \frac{\pi^2}{4} \left(\frac{3}{\hr^3} - \frac{1}{2\hr} +\frac{1}{8}\right) 
+\frac{e^{-\hr}\pi^2}{4}\left(\frac{3}{\hr^3}+\frac{3}
{\hr^2}+\frac{1}{\hr}+\frac{1}{8} \right)\Bigg\}\Bigg] + \mathcal{O}(v)^4.
\label{pot_re_perp_small_v}
\ee

Note that the zero-velocity part of the  potential in Eq.~(\ref{pot_re_para_small_v}) and Eq.~(\ref{pot_re_perp_small_v}) reproduces the previous isotropy potential calculation~\cite{Agotiya:2008ie} and also reproduces the isotropic part of the real part 
of the potential from Refs.~\cite{Thakur:2012eb,Thakur:2013nia}. 

\subsection{Imaginary part of the potential}
 The imaginary part of the potential, $\Im V(\mathbf{r},T,\mathbf{v}) $ from Eq.~(\ref{pot_im_def}) can be written as 
 \be
 \Im V(\mathbf{r},T,\mathbf{v})&=& \int \frac{d^3\mathbf p}{{(2\pi)}^{3/2}}
  (e^{i\mathbf{p} \cdot \mathbf{r}}-1)~V(p)\ \frac{p^2}{2}D_S^{L}({\bf p},u) ~\nn
&=& \int \frac{d^3\mathbf p}{{(2\pi)}^{3/2}}(e^{i\mathbf{p} \cdot 
\mathbf{r}}-1)~\left(\sqrt{(2/\pi)}\ C_F\alpha_s +
    \frac{4\sigma}{\sqrt{2 \pi} p^2}\right)\nn
&\times& \frac{\pi m_D^2 
T(1-v^2)^{3/2}(2+v^2\sin^2\theta)}{2p(1-v^2\sin^2\theta)^{5/2}
\(p^2 + \Pi_R({\bf p},u)\)\(p^2 + \Pi_R^{*}({\bf p},u)\)}.
\ee

We can decompose the total imaginary potential into the Coulombic and string terms as
\be
 \Im V(\mathbf{r},T,\mathbf{v})&=& \Im V_1(\mathbf{r},T,\mathbf{v})+\Im V_2(\mathbf{r},T,\mathbf{v}).
\ee

For the first case, when the dipole is aligned parallel to the direction of the 
relative velocity $ v $ between the dipole and the thermal bath, the Coulombic part of the imaginary potential $\Im V_1(\mathbf{r},T,\mathbf{v})$ can be written as
 \begin{eqnarray}
  \Im V_1(\mathbf{r}\para \mathbf{v},T)&=&\frac{\alpha_s C_F}{2\pi^{2}}\int d^3\mathbf p
      (e^{i\mathbf{p} \cdot \mathbf{r}}-1)\frac{\pi m_D^2 
T(1-v^2)^{3/2}(2+v^2\sin^2\theta)}{2(1-v^2\sin^2\theta)^{5/2}}\nn\
      &\times& 
\frac{1}{p\(p^2 + \Pi_R(p,u)\)(p^2 + \Pi_R^{*}(p,u))}\nn
&=& -2\alpha_s T C_F \int\limits_0^{\pi/2}d\theta\sin\theta 
\frac{(1-v^2)^{3/2}(2+v^2\sin^2\theta)}{2(1-v^2\sin^2\theta)^{5/2}} 
\frac{m_D^2}{\Pi_R^\para(\theta,v) - {\Pi^*}^\para_R(\theta,v)} 
\nn
&+&\Bigg[\Bigg\{ \log x^\para(\theta,v) + \sinh(x^\para(\theta,v)) 
\text{Shi}(x^\para(\theta,v)) -\cosh(x^\para(\theta,v)) 
\text{Chi}(x^\para(\theta,v))\Bigg\} \nn
&-& 
\Bigg\{\log y^\para(\theta,v) + \sinh(y^\para(\theta,v)) 
\text{Shi}(y^\para(\theta,v)) -\cosh(y^\para(\theta,v)) 
\text{Chi}(y^\para(\theta,v))
\Bigg\}\Bigg],\nn
  \end{eqnarray}
  where 
  \be
  x^\para= m_Dr\cos\theta\sqrt{\Pi_R^\para(\theta,v)/m_D^2},\\
  y^\para= m_Dr\cos\theta\sqrt{{\Pi^*}_R^\para(\theta,v)/m_D^2}.
  \ee
and $\text{Shi}(x)$ and $\text{Chi}(x)$ denote the $\uppercase {Mathematica}$ defined functions ${\it SinhIntegral}(x)$ and ${\it CoshIntegral}(x)$, respectively, which can be expressed mathematically as
\be
\text{Shi}(x) = \int\limits_0^x \frac{\sinh t}{t}dt \hspace{2cm} \text{and} 
\hspace{2cm}
\text{Chi}(x) = \gamma_E + \log x + \int\limits_0^x \frac{\cosh t - 1}{t}dt .
\ee

and the string part can be written as
 \begin{eqnarray}
\Im V_2(\mathbf{r}\para \mathbf{v},T)&=&\frac{4\sigma }{(2\pi)^{2}}\int d^3\mathbf p
(e^{i\mathbf{p} \cdot \mathbf{r}}-1)\frac{\pi m_D^2 
T(1-v^2)^{3/2}(2+v^2\sin^2\theta)}{2(1-v^2\sin^2\theta)^{5/2}}\nn
&\times& 
\frac{1}{p^3(p^2 + \Pi_R(p,u))(p^2 + \Pi_R^{*}(p,u))}
\nn
&=& 4\sigma T \int\limits_0^{\pi/2}d\theta\sin\theta 
\frac{(1-v^2)^{3/2}(2+v^2\sin^2\theta)}
{2(1-v^2\sin^2\theta)^{5/2}} \frac{m_D^2}{\Pi_R^\para(\theta,v) - 
{\Pi^*}^\para_R(\theta,v)}\nn
&\times&\Bigg[\frac{\gamma_E + \log x^\para(\theta,v) + \sinh(x^\para(\theta,v)) 
\text{Shi}(x^\para(\theta,v)) -\cosh(x^\para(\theta,v)) 
\text{Chi}(x^\para(\theta,v))}{\Pi_R^\para(\theta,v)} \nn
&-&\frac{\gamma_E + \log y^\para(\theta,v) + \sinh(y^\para(\theta,v)) 
\text{Shi}(y^\para(\theta,v)) -\cosh(y^\para(\theta,v)) 
\text{Chi}(y^\para(\theta,v))}{{\Pi^*}_R^\para(\theta,v)}
\Bigg]\!.\ \ \ \ \ 
  \end{eqnarray}
\begin{figure}[tbh]
\subfigure{
\hspace{-0mm}\includegraphics[width=7.5cm]{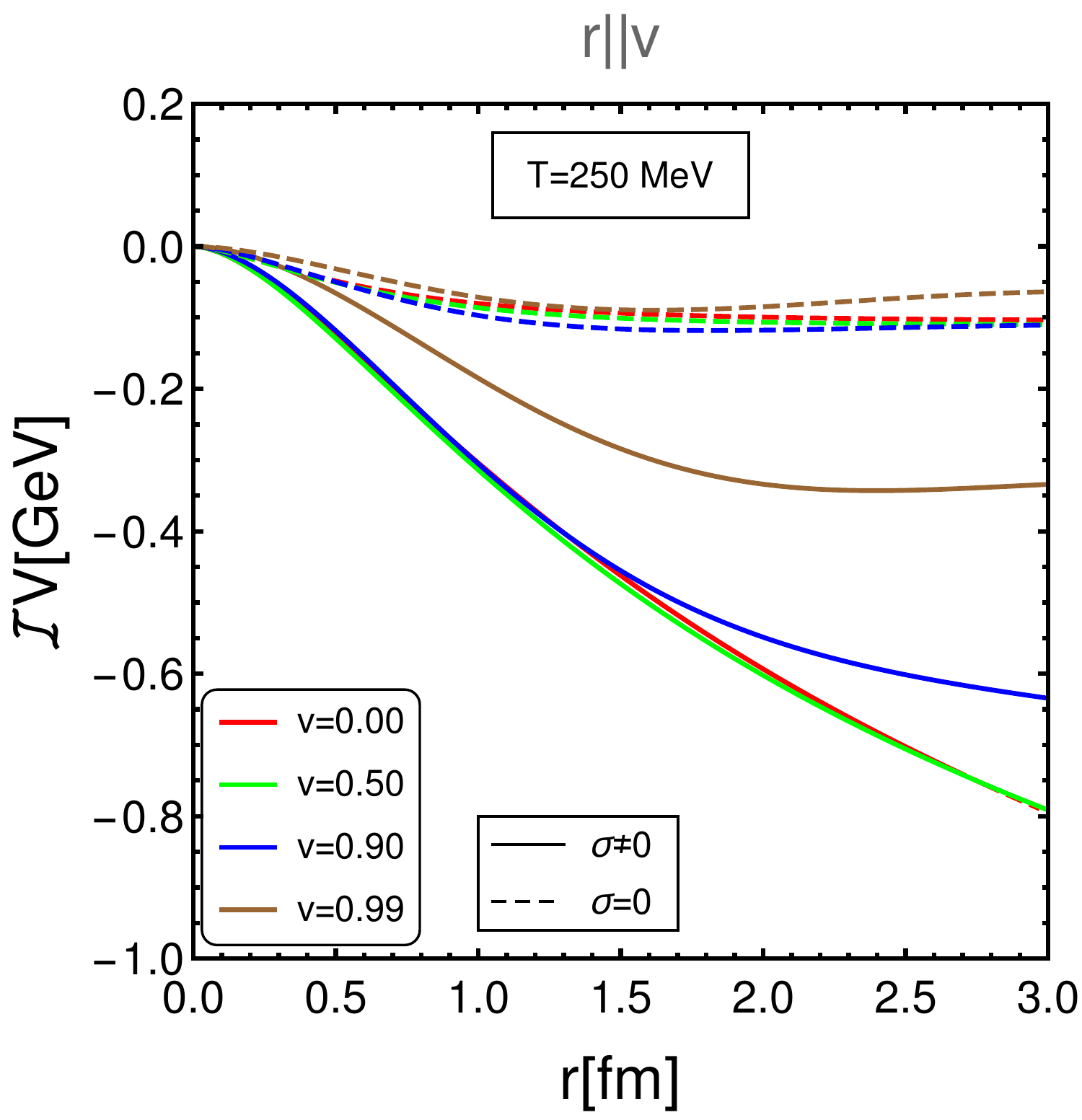}} 
\subfigure{
\includegraphics[width=7.5cm]{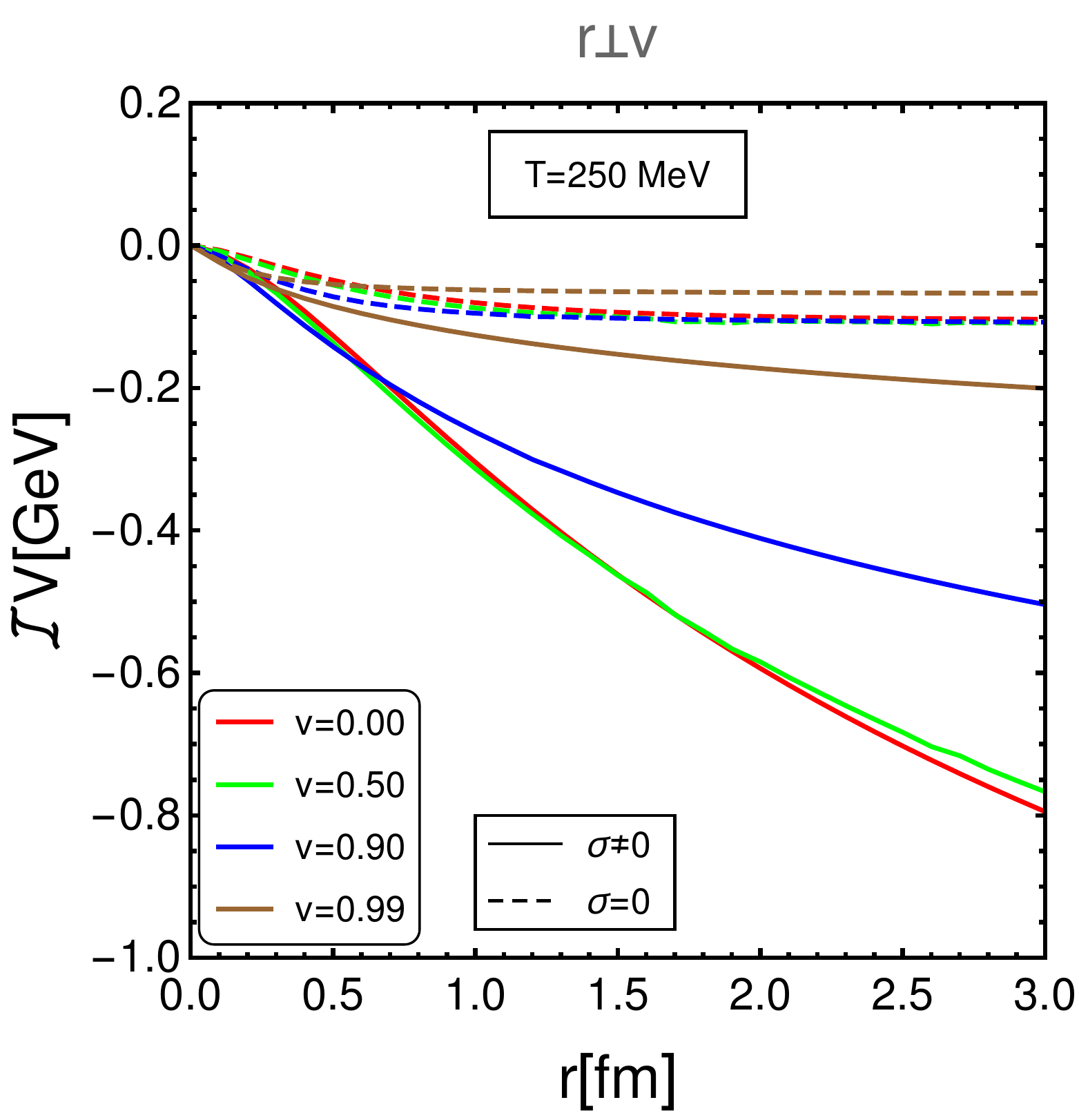}}
\caption{
Variation of the imaginary part of the potential with the separation distance $ r $ between $Q$ 
and $\bar Q$ for various values of velocity at $T=250$ MeV when the dipole axis
is along (left) and perpendicular to (right) the velocity direction.}
\label{pot_im_pr}
\end{figure}

For the second case, when the dipole is aligned perpendicular to the direction of the relative velocity $ v $ between the dipole and the thermal bath, the Coulombic part of the imaginary potential can be written as
\be
\Im V_1(\mathbf{r}\perp \mathbf{v},T) &=& -2\alpha_s T 
C_F\int\limits_0^{\pi/2}d\theta\sin\theta \int\limits_0^{2\pi}\frac{d\phi}{2\pi}
\ \frac{(1-v^2)^{3/2}(2+v^2 - v^2\cos^2(\phi-\beta)\sin^2\theta)}{2\(1-v^2 + 
v^2\sin^2\theta\cos^2(\phi-\beta)\)^{5/2}} 
\frac{m_D^2}{\Pi_R^\perp - {\Pi^*}^\perp} 
\nn
&+&\Bigg[\Bigg\{ \log x^\perp + \sinh(x^\perp) 
\text{Shi}(x^\perp) -\cosh(x^\perp) \text{Chi}(x^\perp)\Bigg\} \nn
&-& 
\Bigg\{\log y^\perp + \sinh(y^\perp) 
\text{Shi}(y^\para(\theta,v)) -\cosh(y^\perp) \text{Chi}(y^\perp)
\Bigg\}\Bigg],\nn
\ee
where
\be
x^\perp= m_Dr\cos\theta\sqrt{\Pi_R^\perp(\theta,\phi,\beta,v)/m_D^2},\\
y^\perp= m_Dr\cos\theta\sqrt{{\Pi^*}_R^\perp(\theta,\phi,\beta,v)/m_D^2}.
\ee
  
Similarly, the string part can be written as

\be
\Im V_2(\mathbf{r}\perp \mathbf{v},T) &=& \frac{4\sigma T }{ m_D^2} 
\int\limits_0^{\pi/2}d\theta\sin\theta \int\limits_0^{2\pi}
\frac{d\phi}{2\pi}\ \frac{(1-v^2)^{3/2}(2+v^2 - 
v^2\cos^2(\phi-\beta)\sin^2\theta)}{2\(1-v^2 + v^2\sin^2\theta
\cos^2(\phi-\beta)\)^{5/2}} \frac{m_D^2}{\Pi_R^\perp - {\Pi^*}^\perp_R}\nn
&&\hspace{.5cm}\times\Bigg[\frac{m_D^2}{\Pi_R^\perp}\Bigg\{\gamma_E + \log 
x^\perp + \sinh(x^\perp) 
\text{Shi}(x^\perp) -\cosh(x^\perp) \text{Chi}(x^\perp)\Bigg\} \nn
&&\hspace{1cm} -\frac{m_D^2}{{\Pi^*}_R^\perp}
\Bigg\{\gamma_E + \log y^\perp + \sinh(y^\perp) 
\text{Shi}(y^\perp) -\cosh(y^\perp) \text{Chi}(y^\perp)\Bigg\}\Bigg].
\ee
\begin{figure}[htb]
\subfigure{
\hspace{-0mm}\includegraphics[width=7.5cm]{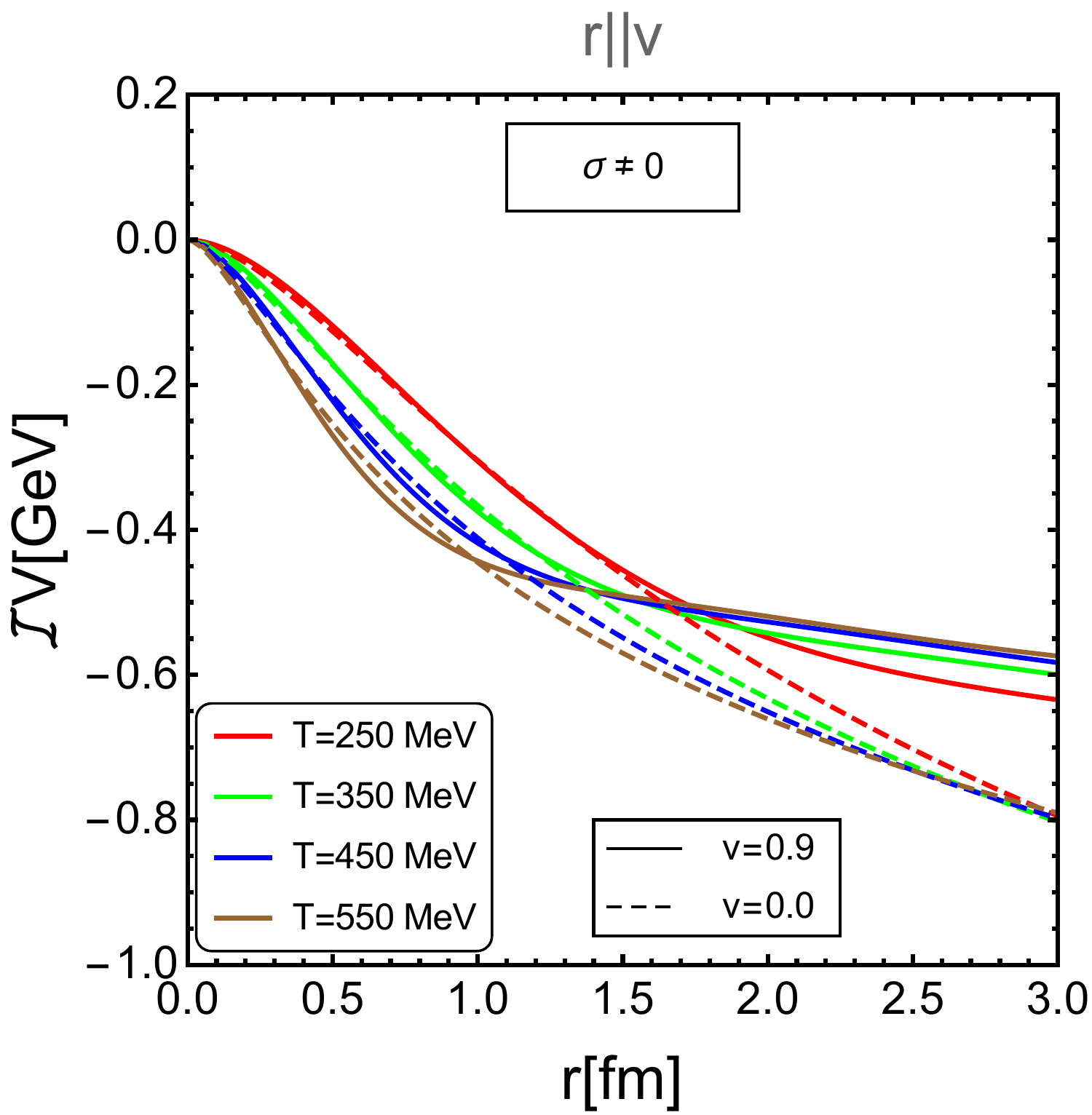}} 
\subfigure{
\includegraphics[width=7.7cm]{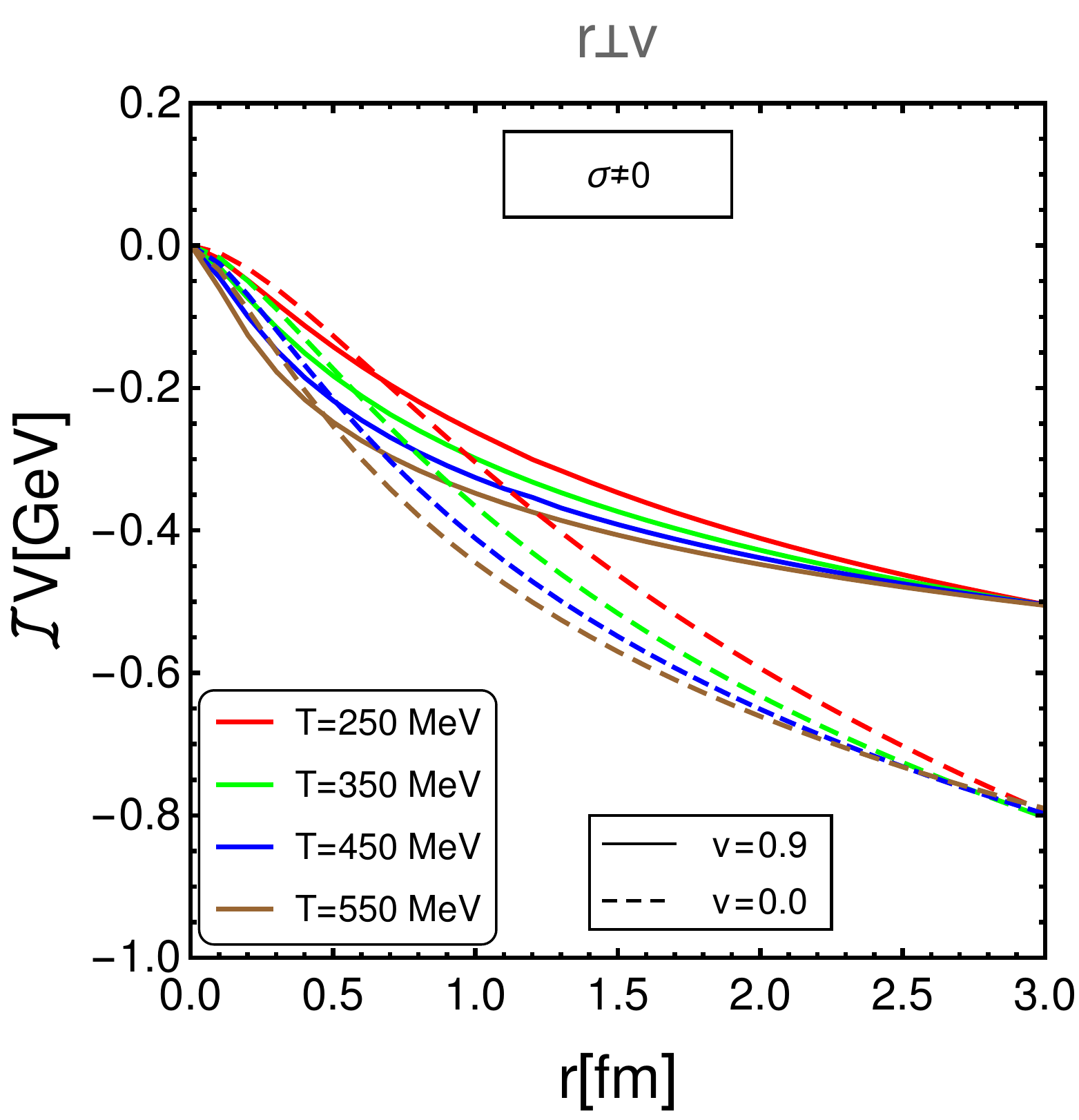}}
\caption{
Variation of the imaginary part of the potential with the separation distance $ r $ between $Q$ and $\bar Q$ for various values of temperature when the dipole axis
is along (left) and perpendicular to (right) the velocity direction.}
\label{pot_im_pp}
\end{figure}

From the Fig.~\ref{pot_im_pr}, we find that the imaginary part of the potential decreases in magnitude at ultrarelativistic velocities and approaches zero. Thus it contributes less to the decay width for both cases at large velocities. The imaginary part 
increases (in magnitude) with the inclusion of the string term ($ \sigma \neq 0 $) as compared to the Coulombic term alone for both the cases. The decrease in magnitude at large velocities is greater for the 
perpendicular case as compared to the parallel case.
The imaginary part vanishes at the origin for any value of $ v $ for both the parallel and perpendicular cases. 
Figure~\ref{pot_im_pp} shows the variation of the imaginary part of the potential with the separation distance $ r $ for various values of temperature, with solid lines for $ v=0.9 $ and dotted lines for $ v=0 $. From the figure we find that the imaginary part of the potential increases in magnitude with an increase in temperature, and hence it
contributes more to the decay width. The increase in magnitude is greater at short distances as compared to large distances. 

Similarly to the real part of the potential in Eq.~(\ref{eq:pot_re_theta}), the imaginary part of the potential at any orientation but at small velocity can be written as
\be
\Im V(\mathbf{r } ,T,\mathbf{v})&=& \frac{1}{2} \Im V(\mathbf{r \para v} ,T)\(1 
+\cos2\theta_{vr}\) + 
\frac{1}{2} \Im V(\mathbf{r \perp v} ,T)\(1 - \cos2\theta_{vr}\) .
\label{eq:pot_im_theta}
\ee

Similarly to the real part of the potential, it is possible to calculate the angular  integration analytically when the relative velocity is small. For small and moderate velocities, the imaginary part of the potential in the parallel alignment case can be
written as
\be
\Im V(\mathbf{r}\para \mathbf{v},T) &\approx& -\alpha_s T C_F\Bigg[1+ \frac{\sinh 
\hr\(\hr\ \text{Shi}\ \hr + \text{Chi}\ \hr\) -
\cosh \hr\(\text{Shi}\ \hr + \hr\  \text{Chi}\ \hr\)}{\hr} 
\nn
&&-\frac{v^2}{192\hr^3}\Bigg\{\frac{\left(35 \pi ^2-432\right) \left(\hr^2+18\right)\hr}{3}\nn
&&+\Big(\text{Chi}(\hr) \sinh (\hr)-\text{Shi}(\hr) 
\cosh (\hr)\Big)\Big(8 \left(\pi ^2-12\right) 
    \hr^4 + 3 \left(35 \pi ^2-432\right) \left(\hr^2+2\right)\Big) \nn
&&\hspace{0cm} - \Big(\text{Chi}(\hr) \cosh (\hr)-\text{Shi}(\hr) \sinh 
(\hr)\Big)\Big(\pi ^2 \hr^4+\left(35 \pi ^2-432\right) 
    \left(\hr^2+6\right)\Big) \hr \Bigg\}\Bigg] \nn
&-& \frac{\sigma T }{m_D^2}\Bigg[(3-2\gamma_E- 2\log\hr) + \frac{\sinh 
\hr\(\hr\ \text{Shi}\ \hr +3\ \text{chi}\ \hr\) -
\cosh \hr\(3\ \text{Shi}\ \hr + \hr\  \text{Chi}\ \hr\)}{\hr}\nn
&& + v^2\Bigg\{\frac{\hr^2 \left(16 \left(\pi ^2-6\right) 
(\log (\hr)+\gamma_E )-35 \pi ^2+112\right) + 90 \left(80-7 \pi ^2\right)}{192 
\hr^2}\nn
&&+\frac{\left(48 \left(2 \hr^4 + 69 \hr^2+150\right)-7 \pi ^2 \left(2 \hr^4+45 
\hr^2+90\right)\right) (\text{Chi}(\hr) \sinh (\hr)-\text{Shi}(\hr) 
\cosh (\hr))}{192 \hr^3}\nn
&&+\frac{\left(\pi ^2 \left(\hr^4+89 \hr^2+630\right)-48 \left(17 
\hr^2+150\right)\right) \hr (\text{Chi}(\hr) \cosh (\hr)-
\text{Shi}(\hr) \sinh (\hr))}{192 \hr^3}\Bigg\}\Bigg] + \mathcal{O}(v)^4 .\nn
\label{Im_pot_para_small_v}
\ee
In addition to the parallel case at small velocity, the imaginary part of the 
potential in the perpendicular alignment case can be
written as
\be
\Im V(\mathbf{r}\perp \mathbf{v},T) &\approx& -\alpha_s T C_F\Bigg[1+ \frac{\sinh 
\hr\(\hr\ \text{Shi}\ \hr + \text{Chi}\ \hr\) -
\cosh \hr\(\text{Shi}\ \hr + \hr\  \text{Chi}\ \hr\)}
{\hr}\nn
&& -\frac{v^2}{192\hr^3}\Bigg\{ \frac{\left(5\(\hr^2-63\)\pi^2 +3888 \right)\hr}{3}\nn
&&- \Big(\text{Chi}(\hr) \cosh (\hr)-\text{Shi}(\hr) 
\sinh (\hr)\Big)\Big(144 \left(\hr^2+9\right)-
5 \pi ^2 \left(2 \hr^2+21\right)\Big) \hr \nn
&& + \Big(\text{Chi}(\hr) \sinh (\hr)-\text{Shi}(\hr) \cosh (\hr)\Big)\Big(48 
\left(\hr^2+3\right) 
\left(\hr^2+9\right)-\pi ^2 \left(\hr^4+45 \hr^2+105\right)\Big) 
\Bigg\}\Bigg] \nn
&-& \frac{\sigma T }{m_D^2}\Bigg[(3-2\gamma_E- 2\log\hr) + \frac{\sinh 
\hr\(\hr\ \text{Shi}\ \hr +3\ \text{Chi}\ \hr\) -
\cosh \hr\(3\ \text{Shi}\ \hr + \hr\  \text{Chi}\ \hr\)}{\hr}\nn
&& + v^2\Bigg\{\frac{\hr^2 \left(16 \left(\pi ^2-6\right) (\log (\hr)+\gamma_E )-35 
\pi ^2+160\right)+45 \left(7 \pi ^2-80\right)}{192 \hr^2}\nn
&&-\frac{\left(48 \left(\hr^4+30 \hr^2+75\right)-\pi ^2 \left(\hr^4+105 
\hr^2+315\right)\right) (\text{Chi}(\hr) \sinh (\hr)-\text{Shi}(\hr) \cosh 
(\hr))}{192 \hr^3}\nn
&& +\frac{\left(48 \left(7 \hr^2+75\right)-\pi ^2 \left(16 
\hr^2+315\right)\right) \hr (\text{Chi}(\hr) \cosh (\hr)-\text{Shi}(\hr) \sinh 
(\hr))}{192 \hr^3} 
\Bigg\}\Bigg]   + \mathcal{O}(v)^4 .\nn
\label{Im_pot_perp_small_v}
\ee

Furthermore, Eqs.~(\ref{Im_pot_para_small_v}) and  (\ref{Im_pot_perp_small_v}) 
can be expanded and written in compact form when the $Q\bar Q$ 
separation is small,
\be
\Im V(\mathbf{r}\para \mathbf{v},T) &\stackrel{\hr\ll 1}{\approx}& -\alpha_s T 
C_F\Bigg[\frac{1}{9}\(4-3\gamma_E -3\log\hr\)\hr^2  + v^2 \(\frac{2}{25}
+\frac{\gamma_E}{10}-\frac{\pi^2}{240} - \frac{\log\hr}{10}\)\hr^2 
+\mathcal{O}(\hr)^4\Bigg] \nn
&-& \frac{2\sigma T}{m_D^2}\Bigg[\frac{\hr^2}{6}+\frac{-107+60\gamma_E+60\log\hr}{3600}\hr^4 \nn
&&+ \frac{v^2}{120}\Bigg\{\(10-\pi^2\)\hr^2 + 
\(\frac{887}{980}-\frac{9\(\gamma_E+\log\hr\)}{7} 
 + \frac{5\pi^2}{168}\)\hr^4\Bigg\} +\mathcal{O}(\hr)^6 \Bigg]  + 
\mathcal{O}(v)^4, \hspace{1cm}
\label{Im_pot_para_small_vr}
\ee
and
\be
\Im V(\mathbf{r}\perp \mathbf{v},T) &\stackrel{\hr\ll 1}{\approx}& -\alpha_s T 
C_F\Bigg[\frac{1}{9}\(4-3\gamma_E -3\log\hr\)\hr^2  + v^2 \(\frac{22}{75}
-\frac{3\gamma_E}{10}-\frac{\pi^2}{720} - \frac{3\log\hr}{10}\)\hr^2 
+\mathcal{O}(\hr)^4\Bigg]  \nn
&-& \frac{2\sigma T}{m_D^2}\Bigg[\frac{\hr^2}{6}+\frac{-107+60\gamma_E+60\log\hr}{3600}\hr^4 \nn
&+&\! \frac{v^2}{360}\Bigg\{\(30-\pi^2\)\hr^2 -\( \frac{1839}{196} - 
\frac{45(\gamma_E+\log\hr)}{7} 
 - \frac{\pi^2}{56}\)\hr^4\Bigg\} +\mathcal{O}(\hr)^6 \Bigg]  + \mathcal{O}(v)^4 .
\hspace{0cm}
\label{Im_pot_perp_small_vr}
\ee

Note that the $v=0$ part of the approximate imaginary part of the potential given in  Eqs.~(\ref{Im_pot_para_small_vr}) and (\ref{Im_pot_perp_small_vr})
matches  the isotropic part of the previously calculated imaginary potential in an anisotropic medium~\cite{Thakur:2013nia}.
\section{Thermal Width}
\label{sec:gamma}
The imaginary part of the potential is a perturbation to the vacuum potential 
and thus provides an estimate for the thermal width ($\Gamma_{Q\bar{Q}}$) for
a resonance state. It can be calculated in a first-order perturbation from
the imaginary part of the potential as
\begin{equation}
\Gamma_{Q\bar{Q}} = - \langle \psi | \Im \, V_{Q\bar{Q}}(\mathbf{r},T,\mathbf{v}) | \psi \rangle,
\label{gamma_def}
\end{equation}
where $ \psi $ is the Coulombic wave function for the ground state and is given 
by
\begin{equation}
\psi({\bf r}) = \frac{1}{\sqrt{\pi a_0^{3}}} e^{-r/a_0},
\label{eq:wavefunction}
\end{equation}
where $ a_0 =2/(C_F m_Q\alpha_s)$ is the Bohr radius of the $Q\bar Q$ system. 
Even though the imaginary potential is not purely Coulombic, the leading
contribution to the potential for the deeply bound $ Q\bar{Q} $
states in a plasma is Coulombic, thus
justifying the use of Coulomb-like wave functions to
determine the width.

Using Eqs.~(\ref{gamma_def}) and (\ref{eq:wavefunction}), the decay width of 
quarkonium in the moving thermal medium can be written as
\be
\Gamma &=&-\frac{1}{\pi a_0^3} \int d^3{\bf r} \,  e^{-2r/a_0}\, 
\Im \, 
V_{Q\bar{Q}}(\mathbf{r},T,\mathbf{v}) 
\label{gamma_exp0}
\ee
As the decay width is defined as the integration over the radial distance $\bf r$,
it does not depend on the orientation of the dipole with respect to the velocity.
For simplicity we take the velocity vector along the $z$ axis. So, the decay width
can be written from Eq.~(\ref{gamma_exp0}) as
\be
\Gamma &=&-\frac{1}{\pi a_0^3} \int_0^{\infty} d^3{\bf r} \,  e^{-2r/a_0}\int 
\frac{d^3\mathbf p}{{(2\pi)}^{3/2}}(e^{i\mathbf{p}
\cdot \mathbf{r}}-1)~\left(\sqrt{(2/\pi)}\ C_F\alpha_s + \frac{4\sigma}{\sqrt{2 
\pi} p^2}\right)\nonumber\\
&&\hspace{1cm}\times \frac{\pi m_D^2 
T(1-v^2)^{3/2}(2+v^2\sin^2\theta)}{2p\(1-v^2\sin^2\theta\)^{5/2}\(p^2 + 
\Pi_R({\bf p},u)\)\(p^2 + \Pi_R^{*}({\bf p},u)\)} \nn
 &=& \Gamma_{1} + \Gamma_{2} ,
\label{gamma_exp}
 \ee
 where $ \Gamma_{1} $ and $ \Gamma_{2} $ are the thermal widths for the 
Coulombic and string parts, respectively.
 The width $ \Gamma_{1} $ for the Coulombic part is
 \begin{eqnarray}
\Gamma_{1} &=&- \frac{\alpha_s m_D^2T C_F}{2\pi^{2} a_0^3} \int d^3{\bf r} \,  
e^{-2r/a_0}  \int d^3\mathbf p
      (e^{i\mathbf{p} \cdot \mathbf{r}}-1)\frac{  
(1-v^2)^{3/2}(2+v^2\sin^2\theta)}{2\(1 - v^2\sin^2\theta\)^{5/2}}\nn
      &&\hspace{1cm} \times \frac{1}{p\(p^2 + \Pi_R({\bf p},u)\)\(p^2 + 
\Pi_R^{*}({\bf p},u)\)}.
\label{gamma1_prth}       
  \end{eqnarray}
We can analytically integrate over  the $ Q\bar{Q} $ separation distance $r$. After 
doing the $r$ integration analytically,
Eq.~(\ref{gamma1_prth}) reduces to
  \begin{eqnarray}
  \Gamma_{1} &=& \alpha_s m_D^2 T C_F 
\int\limits_{-1}^{1}d(\cos\theta)f(v,\theta)\int_{0}^{\infty}\frac{pdp}
  {\(p^2 +\Pi_R(\theta,v)\)\(p^2 + \Pi_R^{*}(\theta,v)\)} 
  \left(1-\frac{1}{\(1+\frac{p^2a_0^2}{4}\)^2}\right),\hspace{.5cm}
  \label{gamma1_pth}
  \end{eqnarray}
where 
  \begin{equation}
  f(v,\theta)=\frac{  
(1-v^2)^{3/2}(2+v^2\sin^2\theta)}{2(1-v^2\sin^2\theta)^{5/2}}
  \end{equation}
and
\be
\Pi_R(\theta,v)&=&\frac{m_{D}^2}{2}\Bigg[ \frac{2-2v^2-v^4\cos^2\theta 
\sin^2\theta}{(1-v^2\sin^2\theta)^2}
-\frac{(2+v^2\sin^2\theta)(1-v^2)v\cos\theta}{2(1-v^2\sin^2\theta)^{5/2}}\nn
&& 
\hspace{4.cm}\times\log\left(\frac{v\cos\theta+\sqrt{1-v^2\sin^2\theta}}{
v\cos\theta-\sqrt{1-v^2\sin^2\theta}}
\right)\Bigg]. %
\ee

We can again perform the momentum integration analytically. After performing the 
integration over the momentum $p$,
Eq.~(\ref{gamma1_pth}) becomes
\begin{eqnarray}
\Gamma_{1}&=&- \frac{\alpha_s m_D^2 T C_F}{2}\int\limits_{-1}^{1}\!d(\cos\theta)f(v,\theta) 
\Bigg[\frac{4 a_0^2}{\left(a_0^2 \Pi_R(\theta,v) -4\right)
\left(a_0^2 \Pi_R^*(\theta,v)-4\right)} - \frac{\log\big(\Pi_R(\theta,v)/\Pi^*_R(\theta,v)\big)}
{\Pi_R(\theta,v) -\Pi^*_R(\theta,v)}
 \nn
 &+& \frac{16}{\Pi_R(\theta,v) -\Pi^*_R(\theta,v)}
 \Bigg\{\!\!\frac{1}{\(a_0^2\Pi_R(\theta,v)-4\)^2}\log \frac{a_0^2 \Pi_R(\theta,v)}{4} 
-\frac{1}{(a_0^2\Pi^*_R(\theta,v)-4)^2}
\log \frac{a_0^2\Pi^*_R(\theta,v)}{4} \Bigg\}
\Bigg]\!.\hspace{0.7cm}
\label{gamma1_th}
 \end{eqnarray}
 
At $v=0$, the above equation reduces to
\be
 \Gamma_{1}(v=0) =  \frac{\alpha_s m_D^2 a_0^2 T C_F}{\(a_0^2 m_D^2 - 
4\)}\left[1-\frac{8}{\(a_0^2 m_D^2 - 4\) }
 + \frac{64}{\(a_0^2 m_D^2 - 4\)^2}\log\frac{m_D a_0}{2}\right].
\ee

 The thermal width $\Gamma_{2} $ for the (nonperturbative) string part is
 \begin{eqnarray}
\Gamma_{2}&=&-\frac{4\sigma m_D^2T }{(2\pi)^{2}a_0^3}\int d^3{\bf r} \, 
e^{-2r/a_0}\int d^3\mathbf p
        (e^{i\mathbf{p} \cdot 
\mathbf{r}}-1)\frac{(1-v^2)^{3/2}(2+v^2\sin^2\theta)}{2(1-v^2\sin^2\theta)^{5/2}
}\nonumber\\
        &\times& 
\frac{1}{p^3\(p^2 + \Pi_R({\bf p},u)\)\(p^2 + \Pi_R^{*}({\bf p},u)\)}
\nn
&=& 2\sigma m_D^2T
\int_{-1}^{1}d(\cos\theta)f(v,\theta)\int_{0}^{\infty}\frac{dp}{p(p^2 
+ \Pi_R(\theta,v))\(p^2 +\Pi_R^{*}(\theta,v)\)}\nonumber\\
  &&  \times \left(1-\frac{1}{\(1+\frac{p^2a_0^2}{4}\)^2}\right).
  \label{gamma2_pth}
   \end{eqnarray}        

After performing the integration over $p$, Eq.~(\ref{gamma2_pth}) becomes  
\begin{eqnarray}
\Gamma_{2}&=& \sigma m_D^2 a_0^2 T 
\int\limits_{-1}^{1}\!d(\cos\theta)f(v,\theta)\Bigg[\frac{a_0^2}{\(a_0^2 
\Pi_R(\theta,v) - 4\)\(a_0^2 \Pi_R^*(\theta,v) - 4\)}
 - \frac{1}{\Pi_R(\theta,v) - \Pi_R^*(\theta,v)}\nn
&\times& \left\{\frac{\(a_0^2 \Pi_R - 8\)}{\(a_0^2 \Pi_R(\theta,v) - 
4\)^2}\log\frac{a_0^2\Pi_R(\theta,v)}{4} - 
\frac{\(a_0^2 \Pi_R^*(\theta,v) - 8\)}{\(a_0^2 \Pi_R^*(\theta,v) - 
4\)^2}\log\frac{a_0^2\Pi_R^*(\theta,v)}{4} \right\}\Bigg].
\label{gamma2_th}
\end{eqnarray} 

At $v=0$, the above equation reduces to
\be
\Gamma_2(v=0) = \frac{4\sigma a_0^2 T}{\(a_0^2 m_D^2 - 4\)^2}\Bigg[4 + 
\frac{a_0^2 m_D^2\(a_0^2 m_D^2 - 12\)}{\(a_0^2 m_D^2 - 4\)}
\log\frac{m_D a_0}{2} \Bigg].
\ee
 \begin{figure}[tbh]
   \subfigure{
   \hspace{-0mm}\includegraphics[width=7.5cm]{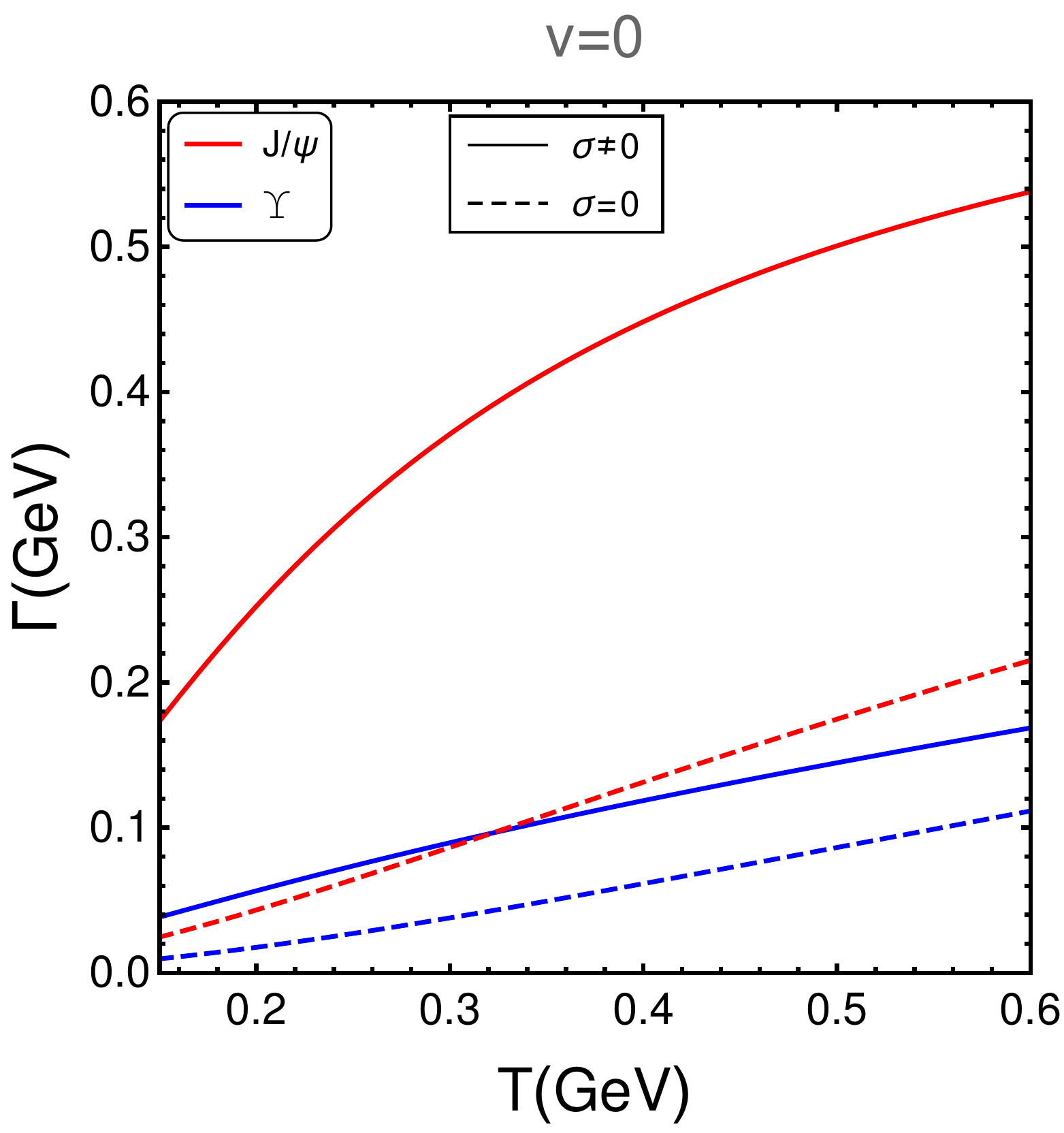}} 
   \subfigure{
   \includegraphics[width=7.5cm]{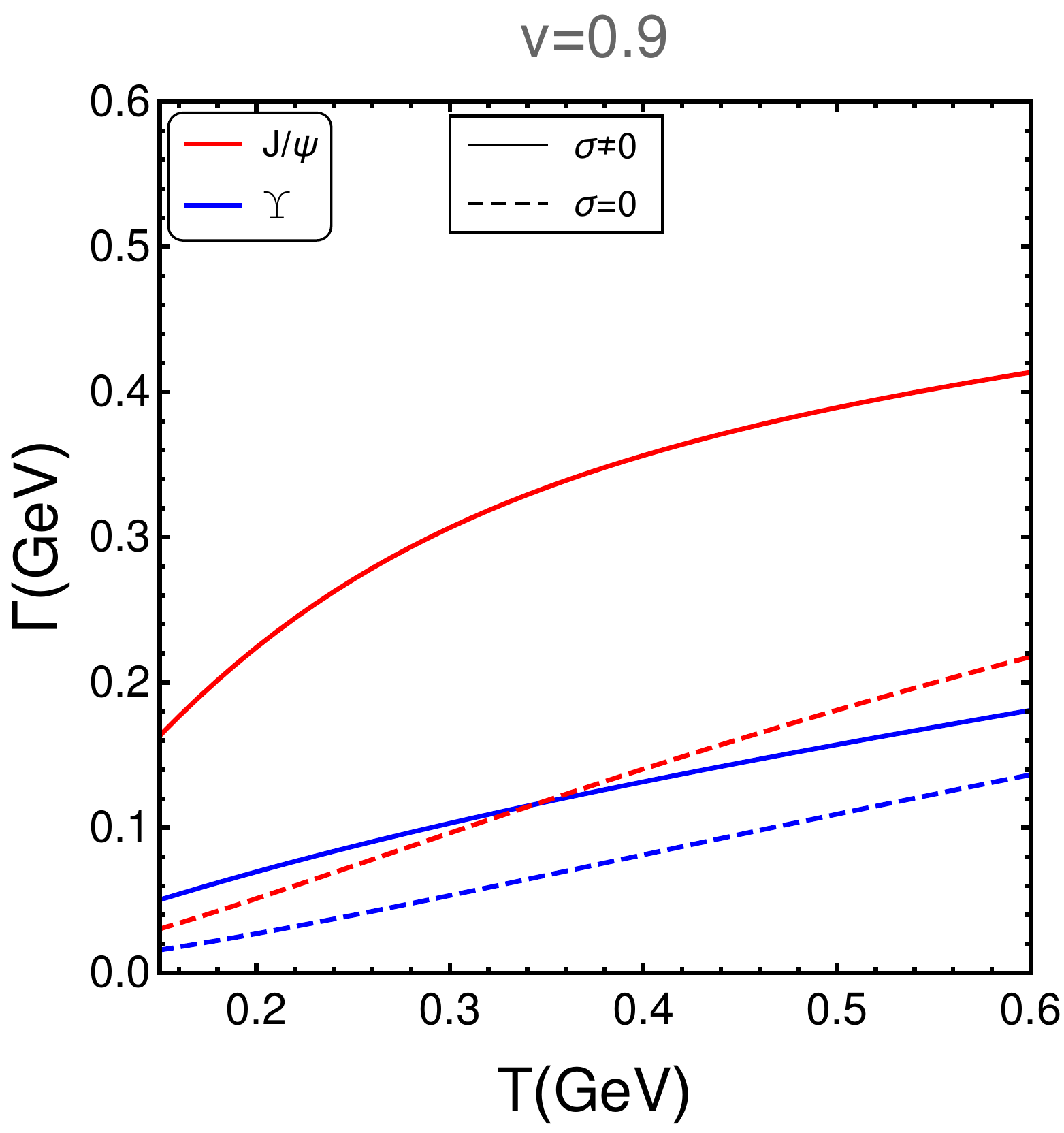}}
   \caption{
   Variation of the decay width with temperature $T$ for $ J/\psi $ and $ \Upsilon $ for $ v=0 $ (left) and $ v=0.9 $ (right). 
}
   \label{gama1}
   \end{figure}
The total decay width, after substituting Eqs.~(\ref{gamma1_th}) and (\ref{gamma2_th}) 
into Eq.~(\ref{gamma_exp}), can be written as
\begin{eqnarray}
\Gamma&=&\!- m_D^2 T \int\limits_{-1}^{1}\! d(\cos\theta)f(v,\theta)\Bigg[ 
\frac{C_F\alpha_s}{2}\left\{\frac{4 a_0^2}
{\left(a_0^2 \Pi_R(\theta,v) -4\right)\left(a_0^2 \Pi_R^*(\theta,v)-4\right)}  
- \frac{\log\(\Pi_R(\theta,v)/\Pi^*_R(\theta,v)\)}{\Pi_R(\theta,v) 
-\Pi^*_R(\theta,v)}\right.
 \nn
 &-&\left. \frac{16}{(\Pi_R(\theta,v) -\Pi^*_R(\theta,v))}
 \left(\frac{1}{\(a_0^2\Pi_R(\theta,v)-4\)^2}\log \frac{a_0^2 \Pi_R(\theta,v)}{4} 
-\frac{1}{(a_0^2\Pi^*_R(\theta,v)-4)^2}
\log \frac{a_0^2\Pi^*_R(\theta,v)}{4} \right)\right\}\nn
&-&\sigma a_0^2\Bigg\{\frac{a_0^2}{\(a_0^2 \Pi_R(\theta,v) - 4\)\(a_0^2 
\Pi_R^*(\theta,v) - 4\)}- \frac{1}{\Pi_R(\theta,v) - \Pi_R^*(\theta,v)}\nn
&\times& \left(\frac{\(a_0^2 \Pi_R - 8\)}{\(a_0^2 \Pi_R(\theta,v) - 
4\)^2}\log\frac{a_0^2\Pi_R(\theta,v)}{4} - 
\frac{\(a_0^2 \Pi_R^*(\theta,v) - 8\)}{\(a_0^2 \Pi_R^*(\theta,v) - 
4\)^2}\log\frac{a_0^2\Pi_R^*(\theta,v)}{4} \right)\Bigg\}\Bigg].
\label{fullwidth}
 \end{eqnarray}
 Note that we compute the momentum integration analytically in all of the cases, 
{\it viz.}, the real part of the potential, the imaginary part of the potential, and the thermal width calculations for both the Coulombic and string terms. We also compute the trivial azimuthal integration analytically in some cases. We  only calculate the nontrivial angular integration numerically. But, in Refs.~\cite{Escobedo:2011ie,Escobedo:2013tca},
the authors have done both the angular and the momentum integration numerically 
for the Coulombic term (except the trivial azimuthal integration.)

We show numerically the variation of the total decay width Eq.~(\ref{fullwidth}) with temperature for the ground states of charmonium and bottomonium in Fig.~\ref{gama1}. We take the charmonium and bottomonium masses as $m_c=1.275$ GeV and $m_b=4.66$ GeV, respectively, from Ref.~\cite{Agashe:2014kda}. We find that the thermal width increases with an increase of the temperature, and the increase in the width with the inclusion of the string term is greater as compared to the earlier result with the Coulombic term alone~\cite{Dumitru:2010id}. The increase in the decay width is less at ultrarelativistic velocities ($ v=0.9$) as compared to the zero-velocity case ($ v=0 $). The width for bottomonium ground state ($ \Upsilon $) is much smaller than the charmonium ground state
($ J/\psi$) because the  bottomonium states are bound more tightly than the charmonium states. 
 \begin{figure}[tbh]
          \subfigure{
          \hspace{-0mm}\includegraphics[width=7.5cm]{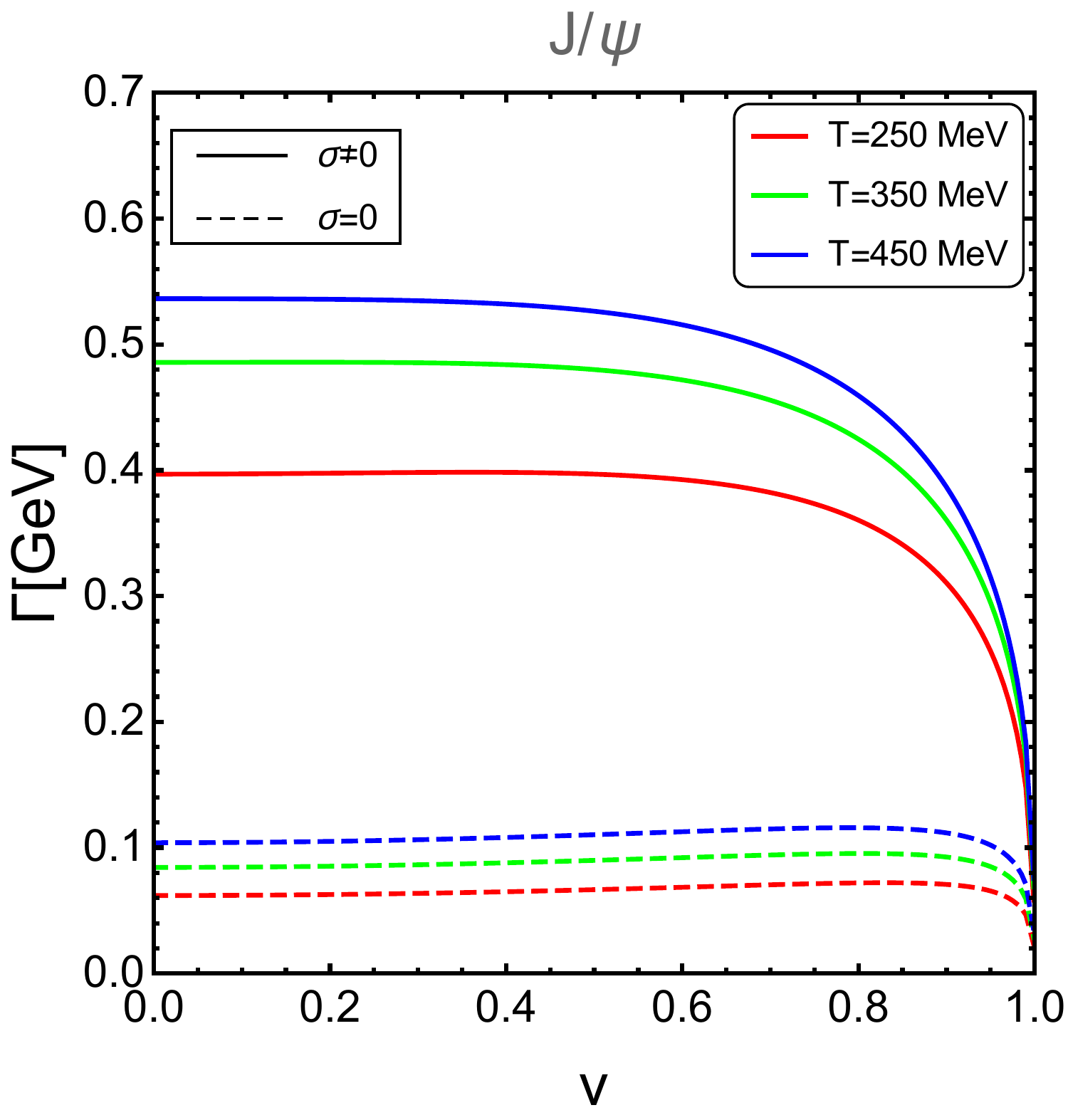}} 
          \subfigure{
          \includegraphics[width=7.5cm]{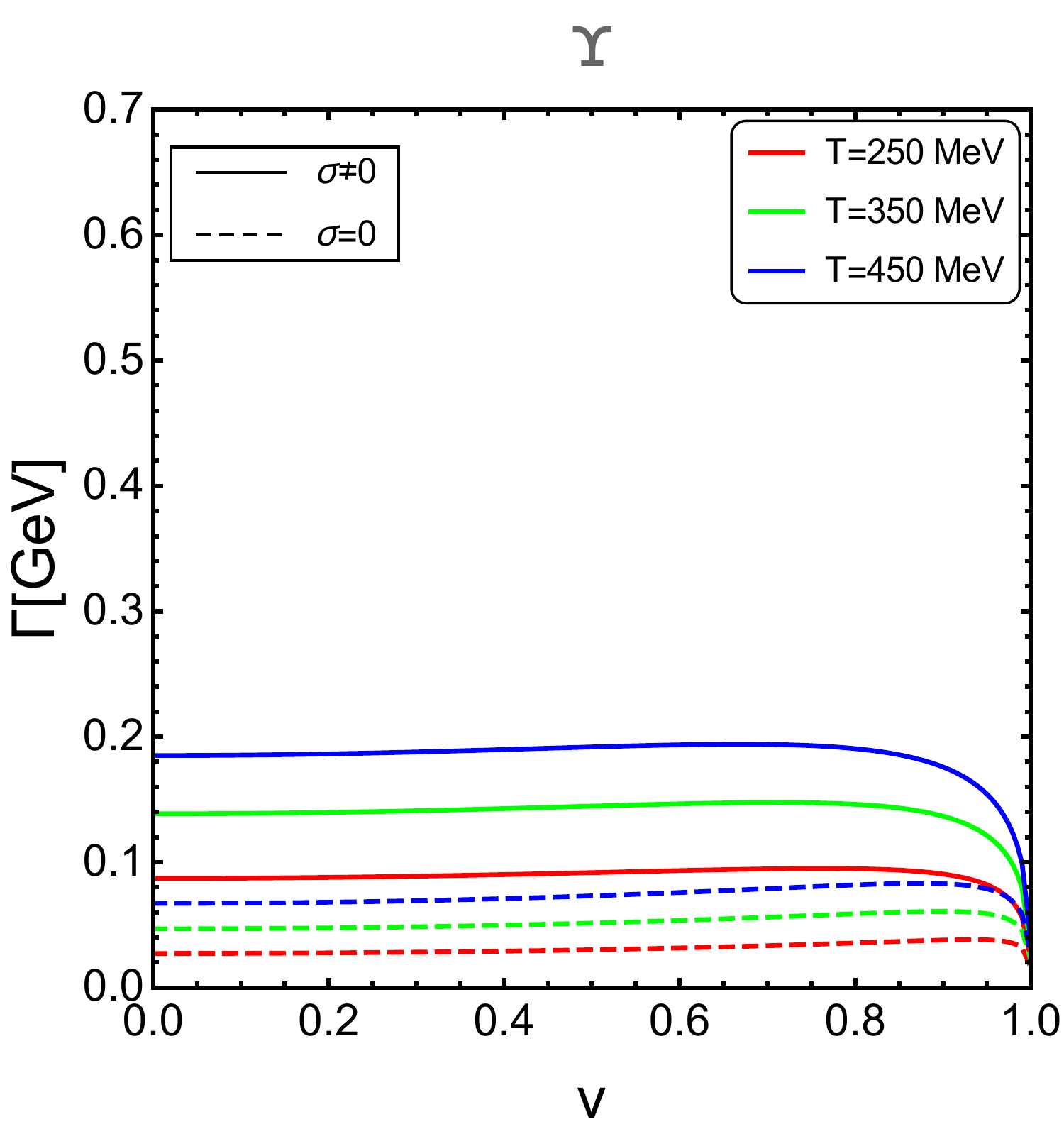}}
          \caption{
         Variation of decay the width for $ J/\psi (1S)$ (left) and $ \Upsilon (1S) 
$ (right) states with velocity $\bf v $ for different values of temperature $T$.}
          \label{gama2}
  \end{figure} 
 Alternatively, we can say that the decay width decreases with an increase in quark masses and dissociation begins at higher temperatures. From Fig.~\ref {gama2}, we find that the width increases with an increase in temperature and decreases at ultrarelativistic velocities.
 
 
\begin{figure}[t]
\subfigure{
\hspace{-0mm}\includegraphics[width=7.5cm]{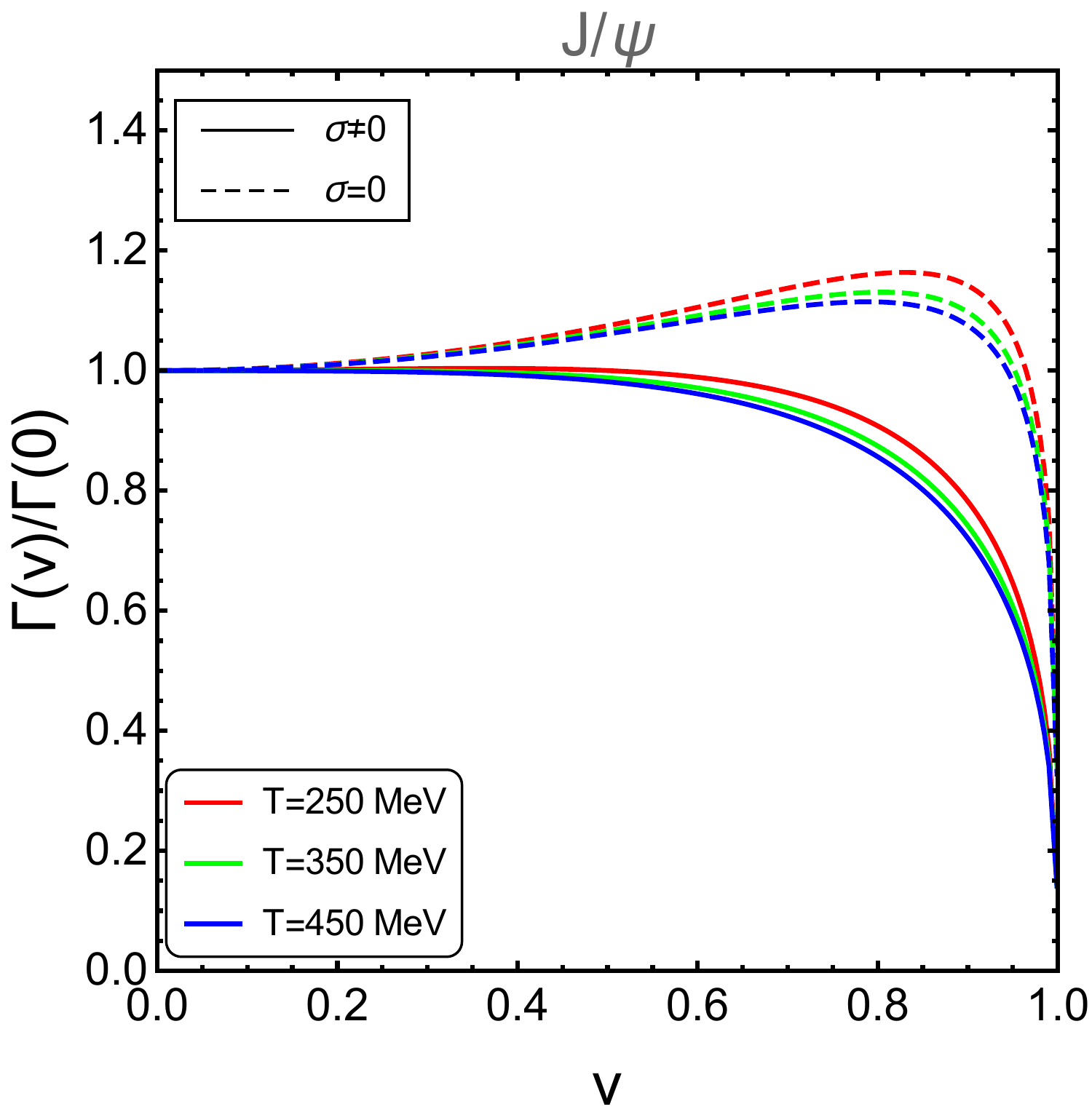}} 
\subfigure{
\includegraphics[width=7.5cm]{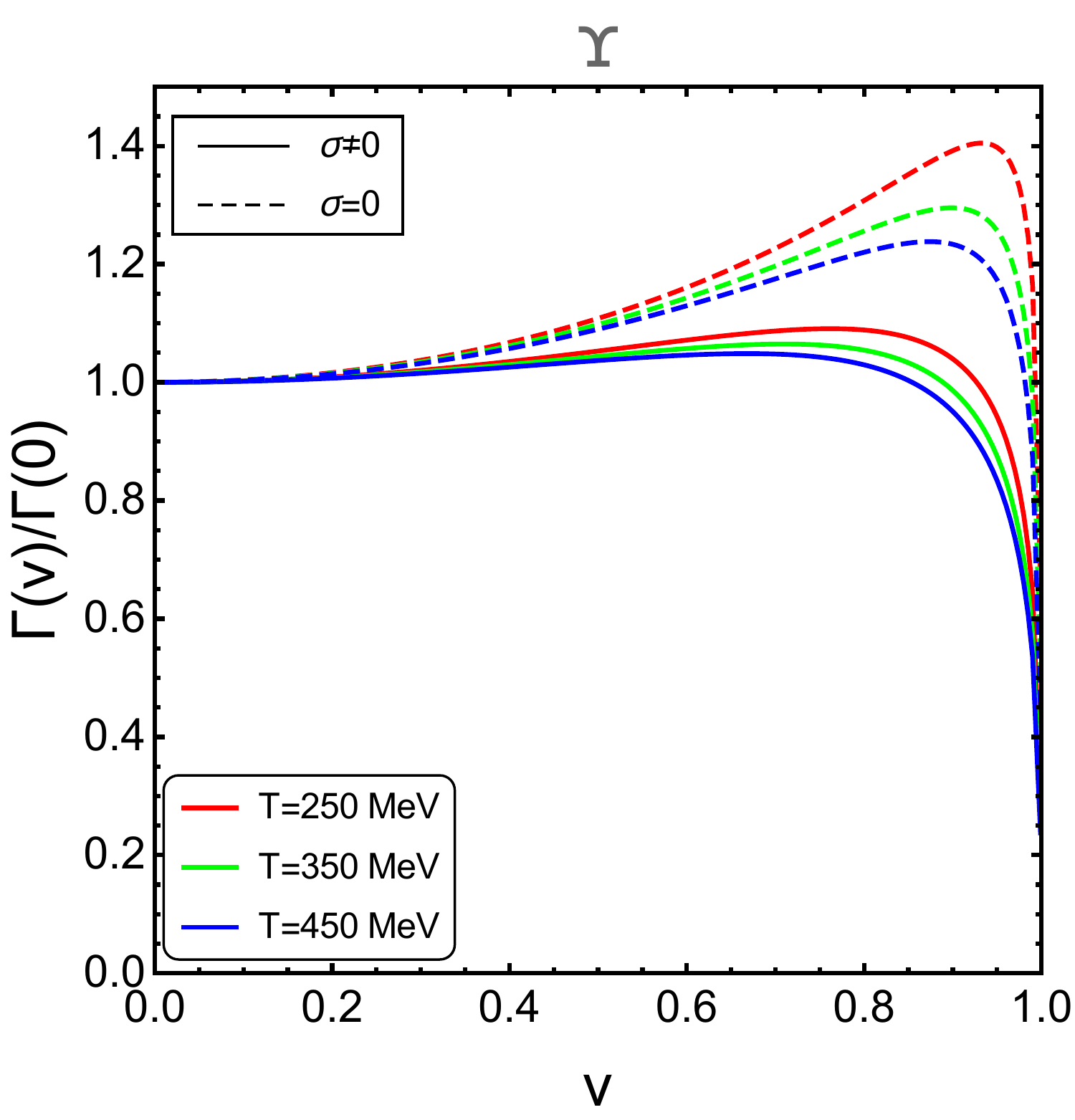}}
\caption{
Variation of the scaled decay width $[ \Gamma(v)/\Gamma(0) ]$ for $J/\psi(1S)$ (left) and $ \Upsilon (1S) $ (right) states  with velocity $\bf v $ 
for different values of temperature $T$.}
\label{gama3}
\end{figure} 
      
  \begin{figure}[tbh]
  \subfigure{
  \hspace{-0mm}\includegraphics[width=7.5cm]{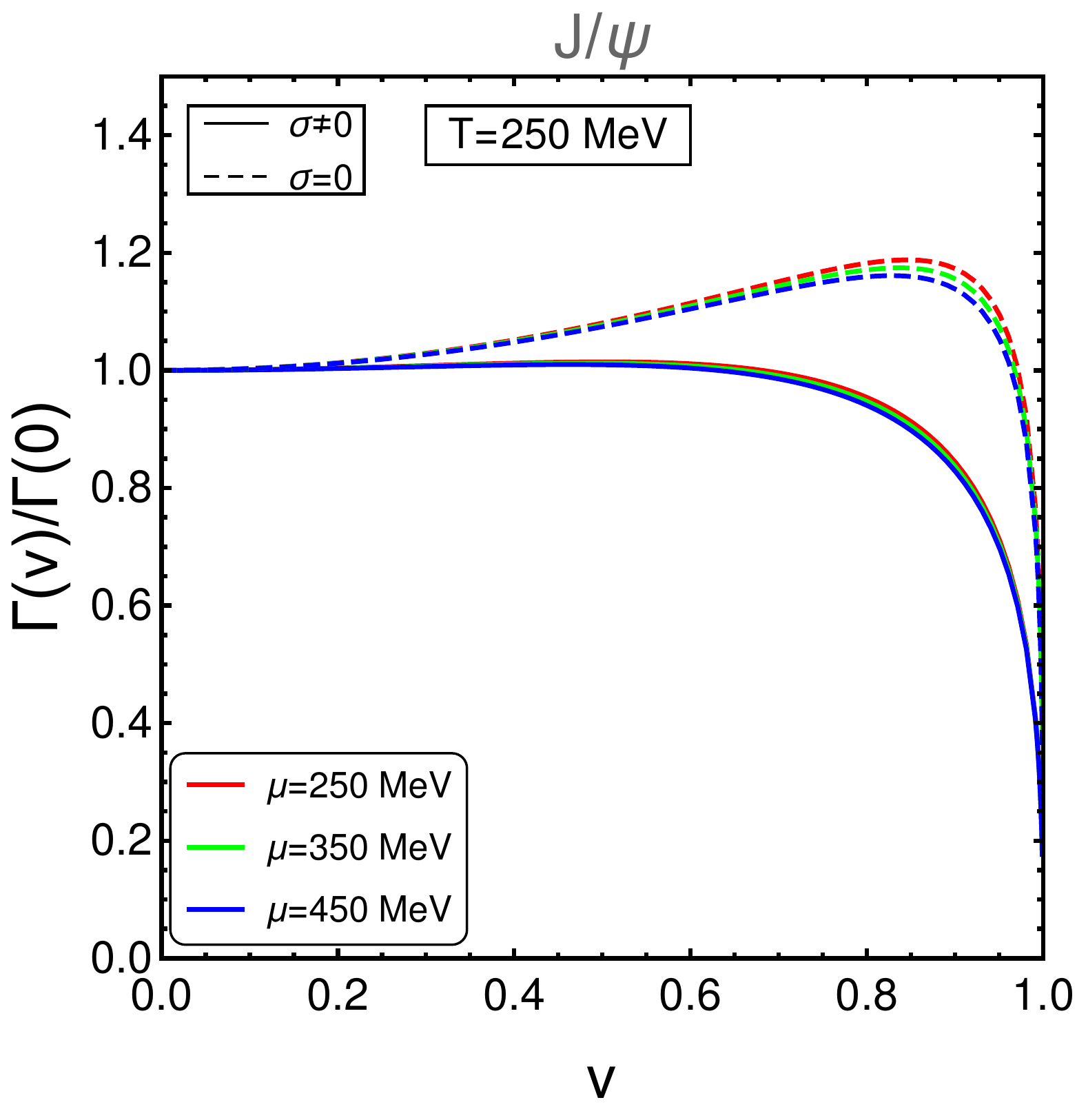}} 
  \subfigure{
  \includegraphics[width=7.5cm]{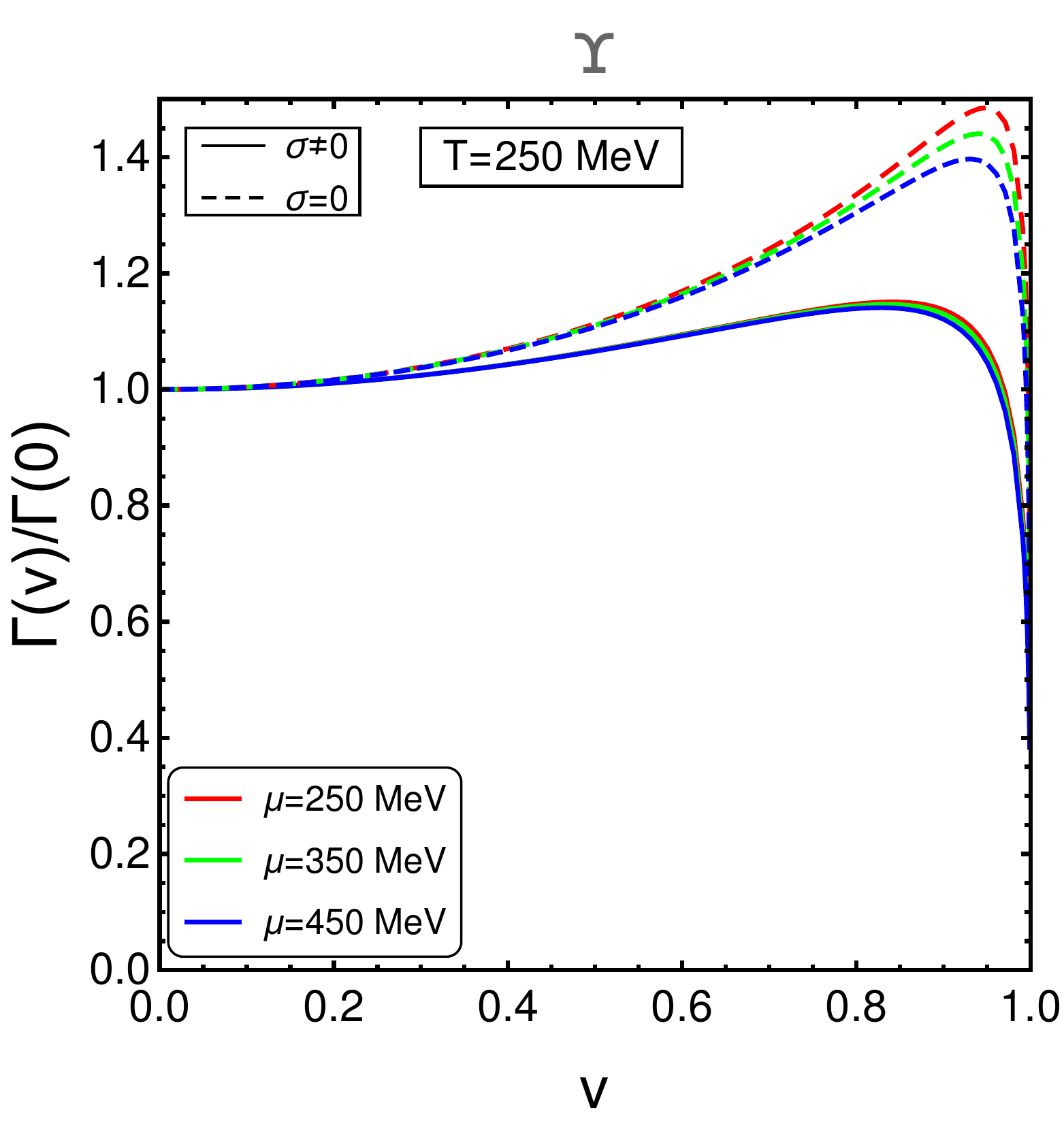}}
  \caption{
  Variation of the scaled decay width $[\Gamma(v)/\Gamma(0) ]$ for $J/\psi(1S)$ (left) and $ \Upsilon (1S) $ (right) states  with velocity $\bf v $ for different values of chemical potential $ \mu $.}
  \label{gama4}
  \end{figure}

Figure~\ref{gama3} shows the variation of the ratio of the total width $[\Gamma(v)/\Gamma(0) ]$ for $ J/\psi (1S)$ and $ \Upsilon (1S) $ states, for three different temperatures, for $ 0 \leq v \lesssim 1 $ with the string term ($\sigma\neq 0  $) and without the string term ($\sigma= 0  $). The ratio of the width decreases for larger values of $ v $ and becomes temperature dependent. This is obvious because, as $ v \rightarrow 1$, Eqs.~(\ref{gamma1_approx1}) and (\ref{gamma2_approx1}) do not hold and other scales 
must be considered ($T_{+} \gg T \gg T_{-} $). The 
decrease in the width ratio at large velocities is greater with the string term as compared to the ratio without string term~\cite{Escobedo:2013tca}, and 
the decrease is greater with an increase in temperature. The reason for the decrease in the decay width is probably related to the fact 
that a moving bound state feels a plasma with a nonisotropic effective temperature ($ T_{eff} $) which we introduced in Sec.~\ref{sec:general_framework}.
 
Figure~\ref{gama4} shows the variation of the ratio of the width $[\Gamma(v)/\Gamma(0) ]$ for $ J/\psi (1S)$ and $ \Upsilon (1S) $ states, for three different values of chemical potential, for $ 0 \leq v \lesssim 1 $ with the string term ($\sigma\neq 0  $) and without the string term ($\sigma= 0  $). The ratio of the width decreases for larger values of $ v $ and becomes much less $ \mu $ dependent. We can say that the width ratio has much less $ \mu $ dependence as compare to the temperature.
\subsection{Approximate decay width} 
We can obtain an approximate solution of Eq.~(\ref{gamma1_pth}) that is valid at moderate velocities and fulfills the relation $1/a_0 \gg m_D $, so that the technique of threshold expansion~\cite{Beneke:1997zp} can be used to work out the integral. Thus, we obtain
 \begin{eqnarray}
 \Gamma_{1} &\approx&  \frac{\alpha_s m_D^2 a_0^2 T 
C_F}{2}\int_{-1}^{1}d(\cos\theta)f(v,\theta)\left[
 \log \left(\frac{2}{m_Da_0}\right)-\frac{1}{4}\right.\nonumber\\
&-&\left. 
\frac{1}{2\(\Pi_R(\theta,v)-\Pi^{*}_R(\theta,v)\)}\Bigg(\Pi_R(\theta,v)\log\frac{\Pi_R(\theta,v)}{m_D^2}-\Pi^{*}_R(\theta,v)\log\frac{\Pi^{*}_R(\theta,v)}{m_D^2}\Bigg)+ 
\mathcal{O}(m_D^2a_0^2)
 \right]\!\!.\ \  
 \label{gamma1_approx1}
 \end{eqnarray}
 
Equation~(\ref{gamma1_approx1}) can be further simplified by taking into account that $\log(\frac{2}{m_Da_0}) $ is logarithmically bigger than the rest of the terms in the parentheses,
 \begin{equation}
 \Gamma_{1} \approx \frac{\alpha_s m_D^2a_0^2 T 
C_F}{\sqrt{1-v^2}}\log\left(\frac{2}{m_Da_0}\right),
 \label{gamma1_approx2}
 \end{equation}
 which exactly matches the approximate decay width in Ref.~\cite{Escobedo:2013tca} without a factor of $2$
 due to the difference in the definition of the decay width in Eq.~(\ref{gamma_def}).
 
Similarly, for the string term, the decay width
that fulfills the relation $ 1/a_0 \gg m_D $ can be approximated at moderate velocities as
 \begin{eqnarray} 
\Gamma_{2}&\approx & \frac{\sigma Ta_0^2m_{D}^2}
{\sqrt{1-v^2}}\left[-\frac{3}{4}a_0^2\log\left(\frac{2}{m_Da_0}\right) +\frac{a_0^2}{8}
+\frac{\log \Pi_R(\theta,v)-
\log \Pi^{*}_R(\theta,v)}{\Pi_R(\theta,v)-\Pi^{*}_R(\theta,v)}\right.\nonumber\\
&+&\frac{3a_0^2}{8\(\Pi_R(\theta,v)-\Pi^{*}_R(\theta,v)\)}\Bigg(\Pi_R(\theta,v)\log\frac{\Pi_R(\theta,v)}{m_D^2}-\Pi^{*}_R(\theta,v)\log\frac{\Pi^{*}_R(\theta,v)}{m_D^2}\Bigg)+ 
\mathcal{O}(m_D^2a_0^2)
 \Bigg]\!.\ \ 
\label{gamma2_approx1}
 \end{eqnarray}
 
Equation~(\ref{gamma2_approx1}) can be further simplified by choosing the leading-order term. But, unlike the Coulombic contribution, the leading-order term in Eq.~(\ref{gamma2_approx1}) is not the logarithmic term; rather, it comes from the third term within parentheses in Eq.~(\ref{gamma2_approx1}). 
Therefore, after neglecting the other terms and keeping the leading-order and next-to-leading-order (logarithmic) terms, we get
 \begin{equation}
 \Gamma_{2} \approx \frac{\sigma Ta_0^4m_{D}^2}{\sqrt{(1-v^2)}}\left[\frac{1}{a_0^2m_D^2} - \frac{3}{4}\log\left(\frac{2}{m_Da_0}\right)\right], 
 \end{equation}
 and the total approximated decay width becomes
 \begin{equation}
\Gamma \approx \frac{a_0^2 m_D^2T}{\sqrt{1-v^2}}\left[\alpha_s C_F\log\left(\frac{2}{m_Da_0}\right) 
+ \frac{\sigma}{m_D^2} \left\{1 -\frac{3}{4}a_0^2m_D^2\log\left(\frac{2}{m_Da_0}\right)\right\}\right]
 \end{equation}
\section{Conclusions and Outlook}
\label{sec:conclusion}
In conclusion, we have studied how the real and imaginary parts of the heavy quark-antiquark potential are modified in the presence of a nonperturbative string
term  when there is relative motion between the dipole and thermal medium.  We have modified the full Cornell potential through the dielectric function in the real-time formalism using the HTL approximation. We have reproduced the previous results for the medium at rest and extended them for nonvanishing velocities
of the thermal medium. We found that the real part of the potential increases with an increase in velocity at short distances and becomes less screened, but it decreases with an increase in velocity at large distances and becomes more screened when the $Q \bar Q$
pairs are aligned along the motion of the medium (parallel case). On the other hand, the potential
decreases with an increase in velocity for the perpendicular case which results in more screening
of the potential. Since  the $ Q\bar{Q} $ potential is effectively more screened in the moving
plasma, it results in earlier dissociation of quarkonium states in a moving medium. The inclusion 
of the string term makes the potential less screened as compared to the Coulombic term alone
for both cases. Combining all of these effects, we expect a stronger binding of a $Q\bar Q$ 
pair in a moving medium in the presence of the string term as compared to the Coulombic term 
alone in a moving medium. We have also shown how the real part of the potential varies at
finite $ \mu $ and found a weak dependence on $ \mu $ up to a chemical potential as large as 
$\mu=550$ MeV. We have observed an oscillatory behavior of the real part of the potential at large
velocities rather than an exponential damping at short distances. This oscillatory behavior 
increases with the inclusion of the string term, which leads to the stabilization of the bound
state. Thus the quarkonium dissociation temperature increases with the inclusion of the string term.

The behavior of the imaginary part of the potential is similar to the one determined for the static medium,
but decreases (in magnitude) with an increase in velocity and increases (in magnitude) with the inclusion
of the string term. As a result, the width of the quarkonium states get narrower at higher relative velocities
and get more broadened in the presence of the string term which plays an important role in the dissociation mechanism.
 The reason for the decrease in the decay width in moving medium is probably related to the fact that 
bound states in a moving medium feel a plasma with a nonisotropic effective temperature. Thus, it is not obvious that the
width of the state in a moving medium should increase or be modified at all when there is a relative velocity between the 
$Q\bar Q$ pair and thermal medium. However, the effective temperature is almost everywhere less than $T$ in the 
ultrarelativistic case due to which the width decreases at $ v\sim1 $ and tends to stabilize the system. In this case the
decay width is dominated by the Landau damping. All of these effects lead to the modification of the quarkonium suppression.
The decay width for the bottomonium ground state is much smaller then the charmonium ground state because bottomonium 
states are tighter than the charmonium state and dissociation begins at a higher temperature. We have also extended our 
calculation at finite $ \mu $ and found a very weak dependence of the decay width on $ \mu $. In future, we would like 
to solve the Schr\"odinger equation with both the real and imaginary parts of the potential to calculate the 
dissociation temperature of quarkonium states by using both the Debye screening and Landau damping mechanism.
It would be interesting to see the dependence of the velocity on the binding energy and dissociation temperature of the 
heavy quarkonium states. 
\section{Acknowledgements}
We thank M. A. Escobedo  and B. K. Patra for useful discussions. We also thank S.~Chakraborty and A.~Bandyopadhyay 
for the useful comments.
N.H. was supported by a Postdoctoral Research Fellowship from the Alexander von Humboldt Foundation. L.T. 
sincerely acknowledges Physical Research Laboratory (PRL), India for the Institute Postdoctoral Research Fellowship.

\end{document}